%% file: 3-j-springer-main.tex
\documentclass{svjour3}    

\usepackage[a4paper, margin=1in]{geometry}
\input{0-imports}

\input{0-definitions}

\newboolean{showWordCount}
\setboolean{showWordCount}{false}

\begin{document}
\ifthenelse{\boolean{showWordCount}}{
  \immediate\write18{texcount -1 -sum -merge -q 2-article.tex output.bbl > 2-article-words.sum }%
  \input{2-article-words.sum} words%

\detailtexcount{2-article}
}

\journalname{}%Education and Information Technologies}

\title{\mytitle}
\titlerunning{\myshorttitle}
%%=============================================================%%
%% Prefix	-> \pfx{Dr}
%% GivenName	-> \fnm{Joergen W.}
%% Particle	-> \spfx{van der} -> surname prefix
%% FamilyName	-> \sur{Ploeg}
%% Suffix	-> \sfx{IV}
%% NatureName	-> \tanm{Poet Laureate} -> Title after name
%% Degrees	-> \dgr{MSc, PhD}
%% \author*[1,2]{\pfx{Dr} \fnm{Joergen W.} \spfx{van der} \sur{Ploeg} \sfx{IV} \tanm{Poet Laureate} 
%%                 \dgr{MSc, PhD}}\email{iauthor@gmail.com}
%%=============================================================%%

\ifthenelse{\boolean{ISanonymous}}
    {\author{Anonymous authors}}
    {\author{\AuthorLH$^{1, 2, *}$ \and
        \AuthorECM$^{2, *}$ \and
        \AuthorSA$^{2}$ \and
        \AuthorFCL$^{2}$ \and
        \AuthorGL$^{2}$ \and
        \AuthorBB$^{3}$ \and
        \AuthorJDZ$^{2}$ \and
        \AuthorFM$^{1, 2}$}}

 \ifthenelse{\boolean{ISanonymous}}
    {\authorrunning{First and second author et al.}}
    {\authorrunning{L. El-Hamamsy  and E.-C. Monnier et al. }} % if too long for running head

 \ifthenelse{\boolean{ISanonymous}}
    {
        \institute{Anonymous affiliations}
    }
    {
        \institute{$^1$ \MOBOTS \\
                   $^2$ \LEARN \\
                   $^3$ \CHILI \\
                   \email{\EmailLH, \EmailECM, \EmailSA, \EmailFCL, \EmailGL, \EmailBB, \EmailJDZ, \EmailFM}}
    }

\date{Received: date / Accepted: date}
% The correct dates will be entered by the editor

\maketitle

\begin{abstract}

\input{1-abstract}

\end{abstract}

% \ifdefined\Anonymous 
% \else
% \section*{Conflict of Interest Statement}
% The authors declare that there are no conflicts of interest

% \section*{Corresponding author}
% Laila El-Hamamsy

% \vfill

% \fi

%\section*{Graphical Abstract}
%\includegraphics[width=\textwidth]{Graphical abstract sustainability article.png}

\input{2-article}

% BibTeX users please use one of
%\bibliographystyle{spbasic}      % basic style, author-year citations
%\bibliographystyle{spmpsci}      % mathematics and physical sciences
%\bibliographystyle{spphys}       % APS-like style for physics
%\bibliography{}   % name your BibTeX data base

%\bibliographystyle{spbasic} 
%\bibliography{cas-refs}

\end{document}

%% file: 0-imports.tex
\usepackage{todonotes}
\usepackage{svg}
\usepackage{array}\newcolumntype{P}[1]{>{\centering\arraybackslash}p{#1}}
\usepackage{ifthen}
\usepackage{xcolor}
\usepackage{natbib}
\usepackage{verbatim}
\usepackage{placeins}
\usepackage{lscape}
\usepackage{multirow}
\usepackage{graphicx}
\usepackage{float}
\usepackage{lipsum}  
\usepackage{lipsum}
\usepackage{todonotes}
\usepackage{longtable}
\usepackage{caption}
\usepackage{booktabs}
\usepackage{enumitem}
\usepackage{tablefootnote}
\usepackage{adjustbox}
\usepackage{afterpage}

\newcommand{\detailtexcount}[1]{%
  \immediate\write18{texcount -merge -sum -q #1.tex output.bbl > #1.wcdetail }%
  \verbatiminput{#1.wcdetail}%
}

\definecolor{lavenderblush}{rgb}{0.93, 0.88, 0.90}%{1.0, 0.94, 0.96}

\colorlet{shadecolor}{lavenderblush}

%% file: 0-definitions.tex
\newcommand{\mytitle}{An Adapted Cascade Model to Scale Primary School Digital Education Curricular Reforms and Teacher Professional Development Programs}

\newcommand{\myshorttitle}{Scaling Primary School Digital Education Curricular Reforms}

\newboolean{ISanonymous}
\setboolean{ISanonymous}{false}
\ifthenelse{\boolean{ISanonymous}}{\newcommand*{\Anonymous}{}}

\newcommand{\KWA}{Educational change}
\newcommand{\KWB}{Scalability}
\newcommand{\KWE}{Curricular Reform}
\newcommand{\KWC}{Professional Development}
\newcommand{\KWD}{Digital Education}
\newcommand{\KWF}{Primary School}

\newcommand{\MOBOTS}{MOBOTS Group, Ecole Polytechnique Fédérale de Lausanne, Lausanne, Switzerland}
\newcommand{\LEARN}{LEARN – Center for Learning Sciences, Ecole Polytechnique Fédérale de Lausanne, Lausanne, Switzerland}

\newcommand{\CHILI}{Computer Human Interaction in Learning and Instruction (CHILI) Laboratory, Ecole Polytechnique Fédérale de Lausanne, Lausanne, Switzerland}

\newcommand{\AuthorLH}{Laila El-Hamamsy}
\newcommand{\AuthorBB}{Barbara Bruno}
\newcommand{\AuthorFM}{Francesco Mondada}
\newcommand{\AuthorJDZ}{Jessica Dehler Zufferey}

\newcommand{\AuthorSA}{Sunny Avry}
\newcommand{\AuthorFCL}{Frédérique Chessel-Lazzarotto}
\newcommand{\AuthorGL}{Grégory Liégeois}

\newcommand{\AuthorECM}{Emilie-Charlotte Monnier}

\newcommand{\EmailLH}{laila.elhamamsy@epfl.ch}
\newcommand{\EmailBB}{barbara.bruno@epfl.ch}
\newcommand{\EmailFM}{francesco.mondada@epfl.ch}
\newcommand{\EmailJDZ}{jessica.dehlerzufferey@epfl.ch}

\newcommand{\EmailSA}{sunny.avry@epfl.ch}
\newcommand{\EmailFCL}{frederique.chessel-lazzarotto@epfl.ch}
\newcommand{\EmailGL}{gregory.liegeois@epfl.ch}

\newcommand{\EmailECM}{emilie-charlotte.monnier@epfl.ch}

%% file: 1-abstract.tex
Many countries struggle to effectively introduce Digital Education (DE) to all K-12 students as they lack adequately trained teachers. While cascade models of in-service teacher-professional development (PD) can rapidly deploy PD-programs through multiple levels of trainers to reach all teachers, they suffer from many limitations and are often ineffective. We therefore propose an adapted cascade model to deploy a primary school DE teacher-PD program throughout an administrative region. The model relies on teacher-trainers who (i) are active teachers in the region, (ii) have a prolonged trainer-PD with experts who piloted the teacher-PD program to acquire adult-trainer and DE-related competences, and (iii) are supported by the experts throughout the deployment. To validate the deployment model we used data from 14 teacher-trainers, the 700 teachers they trained, and 350 teachers trained by experts. The teacher-trainer findings demonstrate that the adapted cascade model effectively addresses most cascade models' limitations. The teacher-related findings further validate the adapted cascade model in terms of perception, motivation and adoption which are at least equivalent to those obtained with the experts. To conclude, the adapted cascade model is an effective means of spreading primary school DE PD-programs at a large scale and can be used in other DE reforms. 

%% file: 2-article.tex
\section{Introduction}
\label{sec:introduction}

Our societies' digital transformations have created considerable challenges for educational systems worldwide. Following decades of discussions, there is a growing consensus regarding the importance of introducing Digital Education\footnote{Digital Education, also referred to as Computing Education, includes Computer Science (CS), Information and Communication Technologies (ICT) and Digital Citizenship} to all students for economic, social, and cultural reasons \citep{webb_computer_2017}. Achieving these objectives inherently relies on widespread curricular reforms to (i) reduce structural barriers by providing access to Digital Education and equal opportunities to all students \citep{wang_diversity_2017, bers_state_2022}, and (ii) ultimately reduce social barriers that are stereotype-related and contribute to under-representation in the field, namely in regards to gender \citep{wang_diversity_2017, sullivan_girls_2016, master_gender_2021}.
Unfortunately, no country, district or school is exempt from the challenges related to implementing widespread reforms \citep{cheung_factors_2012, oecd_curriculum_2020}, particularly since there is no one-size-fits-all solution \citep{clarke_design_2009, coburn_rethinking_2003, lidolf_educational_2020}, reforms tend to affect teachers' usual effectiveness \citep{bransford2005theories}, and sustainable changes in their practices is ``one of the biggest challenges in education'' \citep{hubers_paving_2020}. 
Such challenges are compounded for Digital Education as (i) most teachers were not taught Digital Education in their own formal education and (ii) teachers are generally reluctant to adopt instructional or curricular innovations, specifically those related to technology which are perceived as constantly changing \citep{ertmer_teacher_2010}. It is therefore unsurprising to find that the literature is filled with reports of successes and failures pertaining to the widespread introduction of Digital Education in formal curricula \citep{the_royal_society_shut_2012, balanskat_computer_2015, european_union_digital_2019, bocconi_reviewing_2022}, as well as barriers and facilitators \citep{ertmer_teacher_2010} which require implementing solutions that account for the specificity of the given context. \\

\citet{coburn_rethinking_2003} conceptualises widespread reform into four dimensions: the reform should demonstrate consequential changes in teachers' practices (depth) that are sustained over time (sustainability, \citealp{hubers_paving_2020}) before being spread at a large-scale (spread, which we refer to as scaling in this article), and ultimately internalised by teachers to no longer be perceived as an external reform (shift in reform ownership). It is commonly agreed upon that the first step, affecting changes in teachers' practices, requires implementing effective Professional Development (PD) programs that follow teacher-PD best practices to build teacher capacity and competence
\citep{hayes_cascade_2000, darling-hammond_effective_2017, bocconi_reviewing_2022}. A piloting phase can therefore verify the efficiency of the PD-program and associated resources in getting teachers to change their practices in the short and long-term (sustainability), prior to spreading to all teachers in an administrative region. Nonetheless, it is important to remain attentive to the fact that replicating individual successes ``on a large-scale has proven to be a difficult and vexing issue'' \citep{elmore_getting_1996}. Therefore it is not sufficient to merely pilot an initiative to ensure that the outcomes at a larger scale are effective. There should be an adequate strategy to spread the changes brought about by the reform, and there should be an investigation into whether or not the strategy is effective \citep{wedell_planning_2009} in order to iteratively refine and adapt the process to individual schools' contexts \citep{clarke_design_2009, coburn_rethinking_2003, lidolf_educational_2020}. \\

In this article, we consider the case of a \ifdefined\Anonymous Anonymous Country \else Swiss \fi primary school Digital Education curricular reform that put sustainability and scalability as key issues from the start of the initiative (see section \ref{sec:context}). First, a piloting phase with expert trainers helped validate the PD-program and associated resources, support, and infrastructure using data from teachers from 10 schools. This validation was conducted in multiple stages by considering the PD implementation phase \ifdefined\Anonymous \citep{anonymous_el-hamamsy_computer_2021, anonymous_caneva_technology_2023}\else \citep{el-hamamsy_computer_2021, caneva_technology_2023}\fi, the outcomes 2 years after the PD-program had ended to gain insights into its sustainability \ifdefined\Anonymous \citep{anonymous_el-hamamsy_sustainability_2022}, \else \citep{el-hamamsy_sustainability_2022},\fi and the impact at the student level \ifdefined\Anonymous \citep{anonymous_el-hamamsy_primary_2023}\else \citep{el-hamamsy_primary_2023}\fi. The next step is therefore to consider the spread of the reform to the entire administrative region. In the related work we demonstrate the benefits of a centrally coordinated PD-program for all in-service and pre-service teachers (see section \ref{sec:need_PD_for_all}) to ensure that a sufficient number of teachers are able and willing to teach the discipline.
Furthermore, as the way the reform is implemented is one of the key determinants of its success \citep{tikkanen_lessons_2020}, we explore approaches employed to spread in-service PD-programs at a large-scale (see section \ref{sec:how_PD_for_all}) before proposing a deployment model which attempts to address known limitations (see section \ref{sec:deployment_model}).
The present study therefore centres around the overarching research question ``How can we effectively scale up a pilot teacher-PD program to an entire administrative region, all the while achieving similar outcomes?''. To validate the proposed PD-program deployment model we address the following research questions (see section \ref{sec:eval_methodology}): \\

\begin{enumerate}[label=\textbf{(RQ\arabic*)}, leftmargin=2cm]
    \item What are the benefits and challenges of the proposed deployment model from the teacher-trainers' perspective?
    \item How efficient is the proposed model with respect to the parent pilot program in terms of teachers' perception and adoption of Digital Education pedagogical content?
\end{enumerate}

The analysis employs a concurrent triangulation design (see section \ref{sec:methodology}) which includes: (i) qualitative data from 14 teacher-trainers who need to be trained in both adult-training and Digital Education related skills, (ii) quantitative data from approximately 700 in-service teachers who participated in the first (of three) Digital Education PD deployment phases, and (iii) quantitative data from 350 teachers who participated in the pilot Digital Education PD-program.

\section{Related Work}

\subsection{The need for a centrally coordinated curriculum and Professional Development initiative for all teachers}
\label{sec:need_PD_for_all}

Two main reform implementation strategies are reported in the literature: top-down centralised strategies and bottom-up decentralised strategies \citep{tikkanen_lessons_2020}. Top-down centralised strategies are planned by policy makers and administrators and ensure that external barriers \citep{ertmer_teacher_2010} are addressed (e.g. alignment of decisions with financial support, \citealp{tikkanen_lessons_2020, pietarinen_large-scale_2017}).
Bottom-up decentralised strategies are school-led, and leave schools the autonomy and responsibility of allocating resources and implementing the most adapted solution to their context \citep{oecd_curriculum_2020}. Bottom-up strategies promote educators' ownership of the reform, teacher agency and motivation, but suffer from two main limitations: (i) potentially increasing inequalities between schools \citep{oecd_curriculum_2020}, and (ii) being rarely sustained and scaled \citep{tikkanen_lessons_2020, kawai_contested_2014}. Therefore, if the objective is to promote equity and ensure uniformity across the educational system, a centralised approach should be preferred  \citep{oecd_curriculum_2020, nieveen_balancing_2012}.

Since the ``mechanism by which much innovation in education continues to be introduced is in-service teacher-training'' \citep{hayes_cascade_2000}, there should be a centrally organised reform and teacher-PD that address first order barriers for akk (e.g. financial support, centralised curriculum, pedagogical resources, and teacher PD, \citealp{el-hamamsy_computer_2021}). Such an approach ensures (some) standardisation of the reform and promotes teachers' adoption of novel practices that are aligned with the objectives of the reform \citep{oecd_curriculum_2020, allen_studying_2015, desimone_effects_2002, penuel_what_2007, zehetmeier_sustainability_2009, sullanmaa_curriculum_2019},  all the while avoiding two main issues of decentralised strategies:

\begin{itemize}
    \item An insufficient number of adequately trained teachers to teach the discipline
    \citep{balanskat_computer_2015, oecd_curriculum_2020, bocconi_reviewing_2022}. Indeed, despite a coordinated continuous being ``a critical component in building teacher capacity'' \citep{oecd_curriculum_2020}, decentralised have often been favoured in Digital Education-related contexts.
    The result is that as of 2022, the European commission's review of 25 countries identified that (i) 21 integrate computing in primary school and 22 in secondary school, (ii) but that there are still 18/21 countries at the primary school-level and 21/22 at secondary school-level that lack adequately trained teachers to teach Computing Education \citep{bocconi_reviewing_2022}.

    \item An increase in educators' workload and stress as teachers and school leaders are responsible for implementing the reform, which may inhibit the success of the reform \citep{tikkanen_lessons_2020}). The increase in workload and pressure is even more prominent in the context of Digital Education, and particularly at the primary school-level, as teachers are expected to teach all disciplines. Teachers must therefore must find a way to include the additional content, despite the fact that policy makers do not always adjust the curricular expectations for other disciplines.
\end{itemize}

The most common approach to centrally coordinated in-service PD initiatives is the cascade model which we present in the following section.

\subsection{The cascade model: a means of rapidly scaling up initiatives}
\label{sec:how_PD_for_all}

Cascade models \citep{hayes_cascade_2000, wedell_cascading_2005} rely on a reduced set of experts (level 1 in the cascade) training other trainers (level 2, also referred to as multipliers) to acquire the knowledge they need to deploy the PD-program \citep{roesken-winter_effective_2015}.
These trainers in turn train other groups, a process which is repeated until reaching the lowest levels of the cascade. Cascade models therefore do not require long periods out of service by using ``existing teaching staff as co-trainers'' \citep{gilpin_cascade_1997} and are ``a cost effective means of introducing educational change to large numbers of teachers'' \citep{wedell_cascading_2005} in a short amount of time \citep{engelbrecht_industry-sponsored_2007, ngeze_cascade_nodate}. Cascade models are therefore considered useful in cases where \citep{karalis_cascade_2016} (i) there is a lack of experts to train all recipients, as in the context of K-12 Digital Education and (ii) there is a high number of final recipients, as in the context of widespread curricular reforms.  \\

Unfortunately cascade models, due to both their structure and their implementation, suffer from multiple difficulties which increase their risk of failure \citep{abeysena_cascade_2016}, and therefore have not always been successful \citep{moulakdi_professional_2020, dichaba_does_nodate}. Indeed, cascade models :

\begin{itemize}
    \item are mostly transmissive and do not include feedback between higher and lower levels of the cascade \citep{moulakdi_professional_2020, hayes_cascade_2000, mcdevitt_how_1998, gilpin_cascade_1997}

    \item do not always provide sufficient training \citep{baron_tot_2006} or support \citep{ngeze_cascade_nodate} by experts to enable teacher-trainers to effectively deploy the PD-program.

    \item ``rely on teachers and trainers at different levels to change not only their practices, but also to change their roles while receiving and delivering training'' \citep{abeysena_cascade_2016}. Indeed, level 2 trainers and below are ``both the subjects and agents of change'' \citep{gilpin_cascade_1997} and must acquire adult-training expertise \citep{roesken-winter_effective_2015}.

    \item suffer from content dilution as ``it trickles down'' the cascade \citep{hayes_cascade_2000, wedell_cascading_2005} which may contribute to a decline in the training quality \citep{bax_social_2002, demarle-meusel_educational_2020, dichaba_does_nodate, fiske_elusive_2004}

    \item suffer from issues of alignment with teachers' contexts and needs \citep{wedell_planning_2009, bett_cascade_2016, moulakdi_professional_2020}, notably ``when the trainers do not follow protocols and only provide a portion of the content \citep{bax_social_2002}'' \citep{moulakdi_professional_2020} or when the experts are from another country  \citep{abeysena_cascade_2016}. This is despite the recognised importance of contextual factors \citep{bax_social_2002, wedell_cascading_2005, moulakdi_professional_2020} and adapting the cascaded PD to local realities \citep{bett_cascade_2016}.

\end{itemize}

These challenges combined contribute to \emph{misinterpretations} on the part of teacher-trainers \citep{fiske_elusive_2004, suzuki_effectiveness_nodate}, \emph{lack of confidence} of trainers in lower levels to deliver the training, issues of \emph{legitimacy} \citep{dichaba_does_nodate, ono_case_2010}, and \emph{difficulties planning and managing the training process} \citep{engelbrecht_industry-sponsored_2007, wedell_cascading_2005, bett_cascade_2016}, and may ultimately hinder the success of the deployment.
Therefore, coherently with \citet{karalis_cascade_2016}'s recommendation, cascade models should also include continuous monitoring for quality assurance.
Thus, in this article we propose and investigate an adapted cascade model (see section \ref{sec:deployment_model}) for a widespread Digital Education curricular reform (see section \ref{sec:context}) with an evaluation framework to determine to what extent the initiative is successful and contributes to addressing known limitations (see section \ref{sec:eval_methodology}).

\section{The curricular reform framework and adapted cascade model}

\subsection{Context: a Digital Education curricular reform that considered sustainability and scalability as founding principles from the start}
\label{sec:context}

The study is conducted within the context of a mandatory Digital Education curricular reform project that seeks to introduce Computer Science, Information and Communication Technologies and Digital Citizenship to all K-12 students from 93 schools (approximately 9'000 teachers and 130'000 students). Provided the top-down nature of the initiative, and the challenges of introducing Digital Education into the curriculum, the project sought to address barriers to sustainability and scalability from the start. The key tenets of the curricular reform framework are detailed in Table \ref{tab:curricular_reform_framework} and structured according to \citet{elmore_getting_1996}'s recommendations for effective widespread reforms and can be synthesised as follows:
\begin{itemize}
    \item Having an initiative that is fully funded by the department of education.
    \item Conceiving the reform with key stakeholders to account for the reality of the field and evidence informed best practices.
    \item Providing all the necessary pedagogical and material resources required to teach Digital Education.
    \item Training instructional coaches to support the initiative in-schools and ensure that it the reform is sustained.
    \item Piloting (before deploying at a large scale) a teacher-PD program that follows evidence-based best practices with the associated pedagogical and material resources to help teachers implement Digital Education into their practices.
    \item Having a monitoring framework which includes teacher feedback to iteratively adjust the reform and PD program and address remaining barriers to sustainable change both when piloting and deploying.
\end{itemize}

The effectiveness and limitations of the curricular reform framework were demonstrated through multiple studies conducted in 10 pilot schools at both the teacher- \ifdefined\Anonymous \citep{anonymous_el-hamamsy_computer_2021, el-hamamsy_sustainability_2023} \else \citep{el-hamamsy_computer_2021} \fi  and student-levels \ifdefined\Anonymous \citep{anonymous_el-hamamsy_primary_2023}\else \citep{el-hamamsy_primary_2023}\fi. Therefore, having validated some of the pre-requisites for a sustainable and scalable reform \citep{hubers_paving_2020, coburn_rethinking_2003} the present focus is on effectively spreading the reform to the entire region, all the while addressing known limitations of cascade models in order to achieve similar, or better, outcomes than in the pilot program.

\begin{table}[h!]
\centering
\caption{Curricular reform framework and the link with recommendations for effective widespread reforms}
\label{tab:curricular_reform_framework}

\begin{tabular}{p{4.5cm}p{2cm}p{9cm}}
\toprule
Recommendations for effective widespread reforms & \multicolumn{2}{l}{Corresponding curricular reform framework tenet} \\
\midrule
``Develop strong external normative structures for practice'' and avoid relying on ``the commitment of talented and highly motivated teachers to carry the burden of the reform'' \citep{elmore_getting_1996} & Funding and external support & Fully funded initiative by the department of education to adapt the curriculum, infrastructures, provide pedagogical resources, teacher PD and supporting teachers in the long-term \citep{wedell_planning_2009} \\
 & Conception & Relying on a Research Practice Partnership \citep{coburn_fostering_2021} to conceive and pilot the reform with key stakeholders (experts in Digital Education, pedagogy, teachers and researchers) which is considered by the reform's stakeholders to contribute to the project's quality \ifdefined\Anonymous \citep{anonymous_el-hamamsy_coconstructing_2022}\else \citep{el-hamamsy_coconstructing_2022}\fi, in addition to increasing the likelihood of successfully initiating and sustaining the initiative \citep{oecd_curriculum_2020} \\ \midrule
``Develop organisational structures that intensify and focus rather than dissipate and scatter, intrinsic motivation to engage in challenging practice'' \citep{elmore_getting_1996} & Curriculum and pedagogical resources & Creating a multitude of pedagogical resources (20 activities for computer science, 4 for information and communication technologies and 8 for digital citizenship) that are aligned with the curriculum, \tablefootnote{The up-to-date curriculum is available at \ifdefined\Anonymous (URL removed for peer review) \else \url{https://www.plandetudes.ch/web/guest/education-numerique}\fi, and the 2021-2022 version of the pedagogical resources is available at \ifdefined\Anonymous (URL removed for peer review) \else \url{https://www.vd.ch/fileadmin/user\_upload/accueil/Communique\_presse/decodage.pdf} \fi }
and include a variety of instruction modalities, so that teachers can choose and adapt their teaching accordingly, therefore increasing their agency and the likelihood that they teach the new discipline \ifdefined\Anonymous \citep{anonymous_el-hamamsy_symbiotic_2021}\else \citep{el-hamamsy_symbiotic_2021}\fi. \\
 & In-school support & Training instructional coaches to act as support personnel in the school \ifdefined\Anonymous \citep{anonymous_caneva_technology_2023}\else \citep{caneva_technology_2023}\fi, and drive the communities of practice, a key factor to achieve sustainable changes in teachers' practices following PD programs \citep{lerman_faceface_2008} \\ \midrule
 ``Create structures that promote learning of new practices and incentive systems that support them'', where incentives include ``encouragement and support, access to special knowledge, time to focus on the requirements of the new task, time to observe others doing it'' & Piloting the teacher-PD before deployment & Training teachers to introduce the Digital Education into their practice following evidence-based best practices  (e.g. classroom centred, \citealp{hayes_cascade_2000}, prolonged with time between sessions, balance between theory and practice, isomorphic activities so teachers may first experience the activities as learners before teaching them themselves, access to resources, access to support, ...)  to improve the likelihood of teachers implementing and sustaining the changes \\
 & Teacher feedback & Providing opportunities for teachers to debrief with trainers on their experience through discussion sessions and to provide feedback to researchers through surveys. \\ \midrule
``Create intentional processes for reproduction of successes'' \citep{elmore_getting_1996}, as the successes observed in a specific context may not replicate to others & Monitoring all phases of the reform and teacher-PDs, including piloting and deployment & Monitoring the outcomes of the teacher-PDs to identify and address local barriers that may hinder the initiative using an action research methodology  \citep{casey_unpacking_2018, coghlan_inquiring_2017}, in addition to iteratively refining the curriculum, pedagogical and material resources, and in-school support systems. \\ \bottomrule
\end{tabular}
\end{table}

\pagebreak

\subsection{Proposing an adapted cascade model for effective large-scale deployment}
\label{sec:deployment_model}

\begin{figure}[h!]
    \centering
    \includegraphics[width=0.85\textwidth]{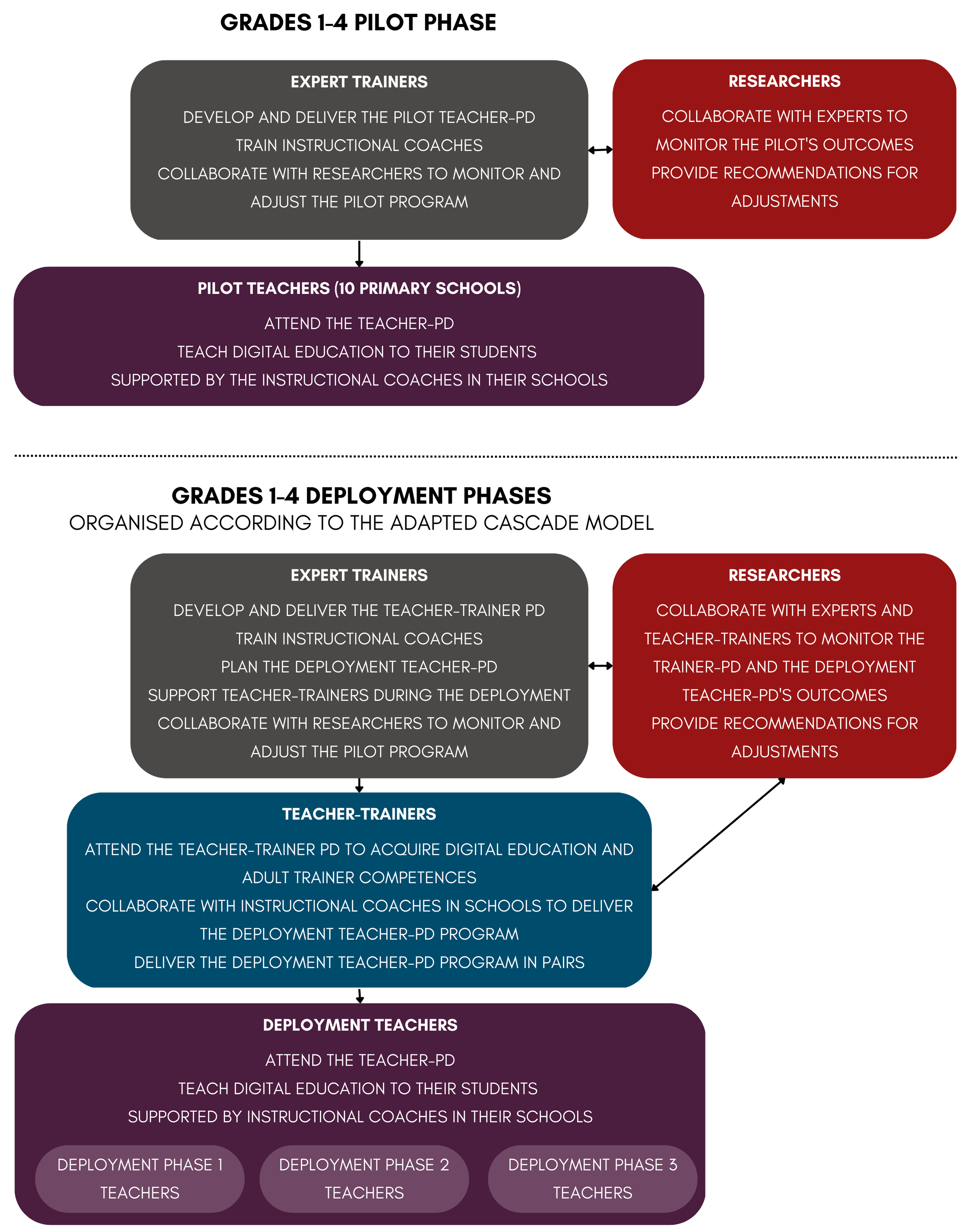}
    \caption{Participants' roles in the adapted cascade model compared to the pilot phase, figure adapted from \ifdefined\Anonymous Anonymous reference\else \citet{monnier_teacher_2023}\fi.}
    \label{fig:roles}
\end{figure}

\begin{figure}[h!]
    \centering
    \includegraphics[width= \textwidth]{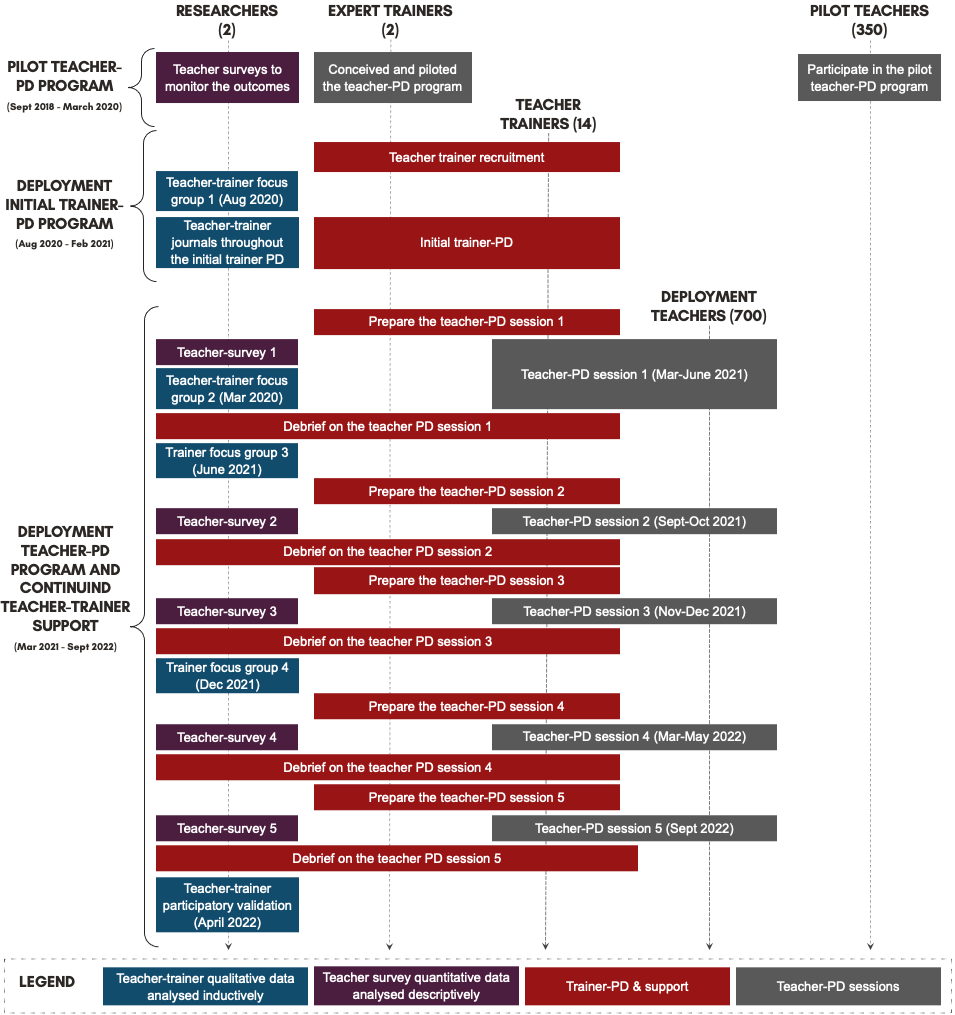}
    \caption{Curricular reform timeline including the pilot program's and the detailed deployment's timelines which includes the trainer- and teacher-PDs and data collections.}
    \label{fig:project_timeline}
\end{figure}

A cascaded deployment model is particularly relevant for Digital Education. As there are few experts and ``a very small segment of the available population could serve as trainers or facilitators'' \citep{glennan_expanding_2000}, these experts cannot reasonably train all teachers in a given administrative region within a short time frame, all the while abiding with teacher-PD best practices. To achieve this objective, we propose and investigate an adapted cascade model to deploy the teacher-PD in three phases to all grades 1-4 teachers in the region (with approximately 1/3 of the teachers per phase).  We describe the model's tenets in the following sections with a synthesis of the main characteristics and the limitations they sought to address in Table \ref{tab:adapted_cascade}. Furthermore, a visual synthesis of the roles and timelines are provided in Fig. \ref{fig:roles} and Fig. \ref{fig:project_timeline}.

\renewcommand{\arraystretch}{1.5}

\begin{table}[h]
\centering
\caption{Characteristics of the adapted cascade model for large scale deployment}
\label{tab:adapted_cascade}

\begin{adjustbox}{width=1.1\textwidth,center}

\begin{tabular}{p{1.5cm}p{6cm}p{4cm}p{4.3cm}}
\toprule
Dimensions & Importance of the dimensions for effective deployment & Adapted cascade model characteristics & Addressing the following limitations of cascade models \\  \midrule

Expert trainers & Expert trainers often from another country which was found to hinder the effectiveness of cascade models in many contexts \citep{bett_cascade_2016, abeysena_cascade_2016}  & Ex-teachers from the region who conceived and piloted the curricular reform and pilot teacher-PD program & Adapting the teacher-PD to the local needs and culture \\

 & Cascade models suffer from content dilution \citep{hayes_cascade_2000, wedell_cascading_2005} which have been found to be directly linked with lower training quality \citep{bax_social_2002, demarle-meusel_educational_2020, dichaba_does_nodate, fiske_elusive_2004} & Directly train all the teacher-trainers who will deploy the teacher-PD (limited cascade model depth) & Limiting content dilution and decline in training quality; promoting teacher-trainer self-efficacy \\

 & Teacher-trainers often struggle with the logistics involved in deploying the PD-program \citep{engelbrecht_industry-sponsored_2007, wedell_cascading_2005, bett_cascade_2016} & Handle the logistics of the deployment & Reducing difficulties planning and managing the teacher-PD \\ \midrule

Teacher-trainers & Teachers are more willing to accept the PD-program and pedagogical content when the trainers are teachers in the region who understand the regional context and have tested the content themselves \ifdefined\Anonymous \citep{anonymous_el-hamamsy_computer_2021} \else \citep{el-hamamsy_computer_2021} \fi & Current teachers from the region & Adapting the teacher-PD to the local needs and culture; promoting legitimacy and credibility \\

& ibid. & Teachers from the level of schooling (ideally) & Aligning with teachers' contexts and needs \\

& Teacher-trainer motivation and retention has been found to be one important limitation of cascade models \citep{mormina_conceptual_2018, burr_faculty_nodate, orfaly_train--trainer_2005} & Motivated towards Digital Education and adult training & Promoting motivation and teacher-trainer retention \\ \midrule

Teacher-trainer PD & Teacher-trainers need to be supported to transition to their new role which includes acquiring the required adult-training and content-specific skills \citep{gilpin_cascade_1997, abeysena_cascade_2016, roesken-winter_effective_2015} & Prolonged initial trainer-PD provided by the experts to acquire the required competences in terms of Digital Education and adult training & Providing sufficient training to acquire the new competences and adapt to their new role; promoting self-efficacy, legitimacy; limiting content dilution \\
& ibid. & Continuing trainer-PD and support by experts throughout the deployment & Providing sufficient support; promoting self-efficacy, legitimacy; limiting content dilution \\ \midrule

Teacher-PD &  Cascade models often suffer from a misalignment between the PD program and teachers' context and needs which is why it is important to adapt the PD to local realities \citep{wedell_planning_2009, bett_cascade_2016, moulakdi_professional_2020} & Deploying the teacher-PD with pairs of teacher-trainers and having these go to the same schools to establish relationships with teachers in the field and ensuring that at least one in each pair is from the cycle & Aligning with teachers' contexts and needs; aligning with PD best practices; providing sufficient support among each other and to teachers; promoting self-efficacy \\

& ibid. & Having teacher-trainers collaborate with instructional coaches in the schools & Aligning with teachers' contexts and needs; providing sufficient support to teacher-trainers  \\

& Insufficient support provided to teachers to help them implement the novel curriculum in the long term contributes to cascade models being ineffective \citep{robinson2002teacher} & Having instructional coaches in schools to support teachers in the implementation of the novel discipline in the short and long term & Aligning with PD best practices; providing sufficient support teachers \\

\midrule

Monitoring & Continuous monitoring of cascade model outputs should be implemented to ensure the quality and effectiveness of the resulting deployment PD program \citep{karalis_cascade_2016} & Having action research as a core component to monitor the outcomes of the deployment & Aligning with PD best practices; validating the effectiveness of the intervention \\

& ibid. & Providing feedback to teacher-trainers and experts to adapt the teacher-trainer PD and the teacher-PD program & Aligning with PD best practices; aligning with teachers' contexts and needs \\

& ibid. & Ensuring that teacher-trainers interact with researchers so that they understand the objectives, methodologies and engage in the process & Aligning with PD best practices; aligning with teachers' contexts and needs \\ \bottomrule

\end{tabular}
\end{adjustbox}

\end{table}

\renewcommand{\arraystretch}{1}

\subsubsection{Envisioned structure of the adapted cascade model}

The deployment model sought to limit the depth of the cascade and have experts in direct contact with all teacher-trainers who would deploy the PD-program in the region.
Indeed, rather than having collaborators or colleagues at the end of the cascade train their colleagues \citep{moulakdi_professional_2020, dichaba_does_nodate}, the PD-program's deployment is carried out by teacher-trainers who were directly trained by experts, similarly to \citet{turner_transfer_2017}'s approach (see Fig. \ref{fig:roles}). \\

\subsubsection{Expert-trainer characteristics}

The expert-trainers in our model are ex-teachers with prior experience in Digital Education and associated reforms. These experts conceived and piloted the PD-program in the region with 10 pilot schools prior to the deployment phase where the PD-program is no longer provided by the experts but by the teacher-trainers in the adapted cascade model.
These experts are therefore well aware of the context in the region and barriers related to the PD-program and its implementation, and can prepare teacher-trainers accordingly. Compared to other cascade models, these experts:
\begin{itemize}
    \item Are in charge of organising the deployment calendar, budget and access to Digital Education resources in the schools
    \item Remain in the field with the teacher-trainers throughout the deployment in order to support and accompany them, and make adjustments to the overall teacher-PD and deployment framework.
\end{itemize}

\afterpage{\clearpage} % flush out floats

\subsubsection{Teacher-trainer recruitment}
\label{sec:recruitement_criteria}

Researchers have proposed several prerequisites for trainers in cascade models, including prior understanding of the training content; having participated in PD on the topics they will be teaching, having attended training activities given by the experts; and being able to deliver the PD with adequate time management \citep{moulakdi_professional_2020, ngeze_cascade_nodate, mormina_conceptual_2018, snowden_achieving_2022}.
However, we consider that most of these characteristics can be addressed in the teacher-trainers' PD (see section \ref{sec:teacher-trainer_PD}). Therefore, we recruited teacher-trainers among the teachers in the region  based on the following characteristics :

\begin{itemize}
    \item Teachers that choose to participate in the program because of their interest in Digital Education and interest to train their colleagues. The teacher-trainer role thus aligned with teacher-trainers' professional goals,  with the objective of improving retention \citep{mormina_conceptual_2018, burr_faculty_nodate, orfaly_train--trainer_2005}.

    \item Teachers who maintain a teaching position in the classroom as their primary activity and reserve at least one day a week for the new teacher-trainer position. This ensures that they are able to test out the content in their classrooms, ``collect examples for further discussion in the course'' \citep{roesken-winter_effective_2015} and increase their credibility and legitimacy \citep{bax_social_2002, moulakdi_professional_2020}. Indeed, one underlying hypothesis is that having teacher-trainers would help teachers that are participating in the PD-program identify with the teacher-trainers and increasing teachers' acceptability of the PD-program.

    \item Teachers that teach in the region, and ideally from the level of schooling that is targeted to ensure that they understand the reality of the field and the context in which they will disseminate the PD program.

    \item Teachers that complement each other in terms of gender, age, prior experience (e.g. adult-training, digital education, art, science, inclusion), and expertise at the given level of schooling in order to complement and support one another both (i) in the field in the teacher-trainer pairs (see section \ref{sec:teacher_PD}) and (ii) as a team.
\end{itemize}

\subsubsection{Teacher-trainers' Professional Development}
\label{sec:teacher-trainer_PD}

The teacher-trainers' PD is led by the experts who piloted the reform and teacher PD-program with the 10 pilot schools. The teacher-trainer PD is directly embedded within the deployment's time frame, with teacher-trainers remaining in contact with the experts until the end of the deployment to the region. We therefore consider the teacher-trainers' PD in two parts: an initial trainer-PD (prior to deploying) and a continuing trainer-PD (while deploying) which sought to adequately prepare and support teachers in their new role (including ensuring that the teacher-trainers were confident in their capacity to train their colleagues, \citealp{orfaly_train--trainer_2005}). \\

The initial trainer PD is a long term active, experiential and reflective PD (as recommended by \citealt{hayes_cascade_2000}, see teacher-trainer initial PD in Fig. \ref{fig:roles}) prior to the first out of three deployment phases for the grades 1-4 teacher-PD sessions (and therefore no longer provided by the expert trainers as in the piloting phase). The objective was to ensure teacher-trainers were prepared to deploy the teacher-PD by :

    (i) acquiring adult training expertise to transition to their new role \citep{gilpin_cascade_1997},

    (ii) acquiring Digital Education-related competences (including those related to computer science concepts, information and communication technologies, and digital citizenship) and

    (iii) creating a cohesive team (with strong interpersonal connections) and community of practice (with members that interact regularly and support each other in their new roles)

The initial trainer-PD therefore lasted approximately 6 months and included at least one weekly day-long session, amounting to a total of 26.5 days (23 in-person, 7 half-days on zoom) of pre-deployment training. This training therefore included multiple instruction formats, as recommended by \citet{roesken-winter_effective_2015} such as theoretical sessions, hands-on sessions where the teacher-trainers experienced first hand the isomorphic pedagogical activities, adult-training workshops, observations of PD sessions delivered by experts and team building sessions.

Specifically to help teacher-trainers reflect on their experience, this initial trainer-PD introduced teacher-trainers to :

    (i) the teacher-PD program's evaluation scheme \ifdefined\Anonymous \citep{anonymous_avry_evaluating_2022} \else \citep{avry_evaluating_2022} \fi to help them understand the indicators for success, and the factors that may influence the outcomes

    (ii) the reflective journal process which help (i) foster self-reflection, (ii) document the teacher-trainers' experience, and (iii) provide feedback to the expert trainers who make adjustments, all with the help of interactions with researchers. \\

The continuing trainer-PD (see teacher-trainer continuing PD, support during teacher-PD and debriefing on teacher-PD in Fig. \ref{fig:roles}) occurs in parallel with the deployment of the teacher-PD program. Indeed, the teacher-trainers continued to be mentored, accompanied and supported by the experts while disseminating the teacher-PD sessions to 700 in-service teachers from 17 schools in the first deployment phase. This approach aligns with recommendations in the literature \citep{wedell_cascading_2005, moulakdi_professional_2020} and ensures that the teacher-PD is deployed in the best conditions possible, while attempting to limit content dilution. To that effect, the accompaniment alternated:

\begin{itemize}
    \item 2-4 day-long preparation sessions with the experts for each teacher-PD session (see teacher-trainer continuing PD in Fig. \ref{fig:roles}) which included giving teacher-trainers access to the pedagogical and training resources they would need \citep{mormina_conceptual_2018}, pre-testing the PD-program's theoretical and practical sessions, and adjusting them based on teacher-trainer's feedback.

    \item 7-9 day-long teacher-PD sessions during which the teacher-trainers supported one another, in addition to the experts' support (see teacher-PD sessions in Fig. \ref{fig:roles}). Furthermore, at the end of each day of teacher-PD, researchers gave the teacher-trainers the teacher-survey results for their session to help them reflect on the day's experience and adjust for upcoming iterations.

    \item 1/2 day of reflexive debriefing (see debriefing on teacher-PD in Fig. \ref{fig:roles}) so teacher-trainers could discuss their individual experiences together and with the experts, given the noted importance of debriefing for teacher-trainers \citep{roesken-winter_effective_2015, ngeze_cascade_nodate}. The researchers would then present the global results based on the PD-program evaluation framework \ifdefined\Anonymous \citep{anonymous_avry_evaluating_2022} \else \citep{avry_evaluating_2022} \fi  which grouped the data for all days and pairs of teacher-trainers'.
\end{itemize}

\subsubsection{Deploying the teacher-PD}
\label{sec:teacher_PD}

The teacher-PD deployment relies on several characteristics to achieve an effective widespread and sustainable curricular reform, including training and support teachers in their environment and therefore within the schools, which we describe below.

The experts who conceived and delivered the pilot PD-program took up the role of the planners in the deployment initiative to avoid teacher-trainers struggling with these elements as observed in other contexts \citep{orfaly_train--trainer_2005, wedell_cascading_2005}. The experts therefore ensured that the necessary Digital Education-specific material and infrastructure had been delivered and set up prior to the PD-session, thus mitigating the risk of teachers being unable to apply what was seen in the training in their own classrooms, an issue raised by \citet{bett_cascade_2016}.

The experts and teacher-trainers then created the pairs of teacher-trainers who disseminated the PD-program together, coherently with best practices \citep{cohen_training_2002}. The objective was to have balanced pairs who complemented each other (see section \ref{sec:recruitement_criteria}) and ensure that delivering 7h of pedagogical content to groups of 15-27 teachers was conducted in good conditions. The pairs of teacher-trainers, once assigned a school and a date, only had to coordinate with the instructional coaches in the schools to ensure that the school was ready for their arrival. The instructional coaches not only to ensure ``both relevant and contextually appropriate training for trainees, and teaching contexts that will support their post-training implementation attempts'' as recommended by \citet{wedell_cascading_2005} but also ensures that teachers have access to short and long-term to teach the discipline. These instructional coaches who therefore play a key role in the deployment framework were all trained by the same experts who trained the teacher-trainers, and can therefore be viewed as a parallel second level in the cascade model, an element which facilitates the implementation and coherence of the reform.

\section{Methodology}
\label{sec:methodology}

\subsection{Participants and Data Collection}
\label{sec:eval_methodology}

To investigate the efficiency of our adapted cascade model (see section \ref{sec:deployment_model}), it is important to include the perspective of the two key stakeholders: the teacher-trainers to ensure that the deployment model has adequately prepared them to train teachers and addressed known limitations of cascade models (RQ1), and teachers to evaluate the impact and efficiency of the deployment model (RQ2). To that effect the study employs a concurrent triangulation design which includes qualitative teacher-trainer data (see section \ref{sec:teacher_trainer_data}) and quantitative teacher data (see section \ref{sec:teacher_surveys}). The timeline for each of these data collections can be seen in Fig. \ref{fig:project_timeline}.

\subsubsection{Teacher-trainers and qualitative journals and focus groups}
\label{sec:teacher_trainer_data}

Following a call to all teachers in the region, 14 teacher-trainers were trained and deployed the teacher-PD\footnote{Please note that 15 teacher-trainers (10 women and 5 men) were recruited to deploy the grade 1-4 (students aged 5-9 years old) teacher PD-program. However, one chose not to continue on as teacher-trainers after the initial trainer-PD and the first teacher-PD session due to the additional load provided by this new role. The remaining 14 all participated in the teacher-trainer data collections. Furthermore, at the time where the article was written, of the 14 just 10 remain as two stopped for personal reasons (moving, receiving a job offer) and two because of the project. Therefore, out of the initial 15, three stopped for reasons which are due to the deployment model, and two for personal reasons. The retention rate is therefore of 67\% overall, and 77\% when accounting for those who could not stay on for personal reasons. It is therefore essential to consider recruiting a larger number of teacher-trainers at the start to ensure that a sufficient number carry on over time.}.
These teacher-trainers are mainly women (10/14), have prior experience as instructional coaches in the Digital Education curricular reform (10/14), and half of them teach in the target grades (i.e. grades 1-4, ages 4-9). Throughout the first deployment phase, and between August 2020 and September 2022, the teacher-trainers participated in the deployments' evaluation scheme which included two main sources of data: data from the their reflexive journals and data from four focus groups. Please note that there was also a participatory validation session with the teacher-trainers which confirmed that the researchers' conclusions matched the participants' experience but that we do not present in detail in this article to leave space for journals, focus groups, as well as the teacher survey data from the pilot program and deployment phase (see section \ref{sec:teacher_surveys}).

\paragraph{Teacher-trainer reflexive journals to gain insight into teacher-trainers' experience throughout their PD and deployment.}

Part of the teacher-trainers' PD program included reflexive journals. In addition to contributing to their reflection on their experience as recommended by \citet{roesken-winter_effective_2015}, they provided a means for the teacher-trainers to provide feedback on their PD. This feedback was analysed weekly by a researcher who then debriefed with the experts and the teacher-trainers in order to iteratively refine the trainer-PD program. As the researchers were the only ones to have access to the journals and the trainers' identity, the feedback to the experts was anonymous, which meant that the teacher-trainers could express themselves freely on the topics they felt were relevant without fear of judgement. The objective was to get unfiltered insight into their experience, in the hopes of minimising the risks of social desirability biases that are common in qualitative settings \citep{bergen_everything_2020}.

\paragraph{Teacher-trainer focus groups to evaluate the characteristics of the adapted cascade model.}

To complement the reflexive journals and address more specific topics that were of interest to the researchers and experts, four focus groups were organised with the teacher-trainers. The advantage of this approach over individual interviews is to encourage discussions between members of the group and potentially elicit insight into elements that might not be expressed in a one-to-one interview setting \citep{ho_focus_2006}. Although focus groups run the risk of social desirability biases, two organisational elements attempted to mitigate this phenomenon. The first is that researchers, and not the experts (who were not privy to the discussions during the sessions), moderated of the focus groups. The second is that the teacher-trainers were already familiar with one another and therefore already trusted each other to a certain extent, which makes it more likely that they engage in free speech \citep{kitzinger_methodology_1994}. The teacher-trainers were split into two groups for the focus groups (except for the third for organisational reasons) to ensure they had enough time to express themselves. Each focus group lasted between 45 and 75 minutes, with the topics addressed per focus group being provided in Table \ref{tab:focus_groups}.

\subsubsection{Teachers and quantitative surveys}
\label{sec:teacher_surveys}

The teacher-data for the deployment phase was collected through surveys at the end of each of the teacher-PD session to gain insight into their perception of the PD-program, pedagogical content, and adoption (see Table \ref{tab:teacher_surveys}), and align with the evaluation framework used in the pilot program \ifdefined\Anonymous \citep{anonymous_el-hamamsy_computer_2021, anonymous_avry_evaluating_2022} \else \citep{el-hamamsy_computer_2021, avry_evaluating_2022}\fi. Approximately 700 teachers (95\% women) participated in the deployment phase and we obtained between 550 and 640 responses for each PD-session (see Table \ref{tab:deployment_demographics}). This data is compared (when comparable metrics are available) with the data from the 350 teachers (see \ifdefined\Anonymous \citet{anonymous_el-hamamsy_computer_2021, anonymous_el-hamamsy_sustainability_2022}\else \citet{el-hamamsy_computer_2021, el-hamamsy_sustainability_2022}\fi) who were trained by the experts during the piloting phase (see Table \ref{tab:pilot_demographics}). Please note that (i) the pilot phase was initially planned over 8 sessions, while the deployment was shortened to 6 after adjustments based on the evaluation of the pilot program and that (ii) data could not be collected during the 6th and final deployment PD-session.

\begin{table}[h]
\centering
\caption{Teacher-trainer semi-structured focus groups}
\label{tab:focus_groups}
\begin{tabular}{cp{2cm}p{1.7cm}p{10cm}}
\toprule
Focus & When & Topic & Objective \\
group & \\\midrule
1 & During the teacher- & Reflexive journals & Explore the utility of the journals and how their use may be improved to align with the teacher-trainers' needs \\
 & trainers' initial PD (December &  &  1. Can you describe the personal experience you had with reflexive journal assignment? \\
 & 2020)  &  &  2. How much time per week did you spend on the reflexive journal? How did you use your reflexive journal? For which purpose?  \\
 &  &  &  3. Did you find the expert feedback useful and why?  \\
 &  &  &  4. How did the reflexive journal make you more comfortable with the tablet?  \\
 &  &  &  5. How do you feel about the upcoming teachers’ session? \\ \midrule

2 & While deploying the first & Experience in the field & Explore all the obstacles and facilitators related to their first experience as teacher-trainers in the field \\
 & set of teacher-PD sessions  &  &  1. Can you describe your experience during the first teacher-trainer sessions? \\
 & (March 2021) &  & 2. Can you describe your experience during the first teacher-PD sessions you led in the field?  \\
 &  &  &  3. What went well? what went less well? what needs to be adapted?\\ \midrule

3 & After deploying the first set  & Experience in the field & Explore all the obstacles and facilitators related to their first experience as teacher-trainers in the field \\
 & of teacher-PD sessions (May 2021)  &  &  1. Can you describe the personal experience you had since our last focus group regarding the following themes which were evoked in the previous focus group? (Content / Context / Timing / Relevance and audience / Relationships  / Teacher reactions / The teacher-trainer PD \\
 & &  &  2. Did you experience something new in the field since the last focus group? \\
 &  &  &  3. Do you have any other comments? \\
 \midrule

4 & After deploying the second  & Reflection on the cascade  & Identify whether the known limits of cascade models are addressed by the present deployment model with the following questions: \\
 & / third set of teacher-PD  & model and its limits &  1. Legitimacy: How do you feel regarding your legitimacy to train your teachers? \\
 & sessions (December 2021) &  &  2. Dilution of content: Do you consider that you give the training sessions the same way over the course of the PD-sessions? What do you think is the impact for teachers? \\
 &  &  &  3. Dilution of content / self-efficacy: How do you think teachers learn as well with you as they would with the experts? \\
 &  &  &  4. Complementarity between teacher-trainers and instructional coaches:  What kind of relationships do you have with the instructional coaches? Do you think this helps you to fulfil your mission as a teacher-trainer? \\
 &  &  &  5. Support between teacher-trainers: How do you support each other as teacher-trainers? What do they bring to you? What do you bring to them? \\
 &  &  &  6. Controllability and sense of ownership: Do you feel you have a say in your PD and in the teachers' PD? How is your feedback on the PD program taken into account? \\
 &  &  &  7. Motivation and retention: How do you perceive your future as a teacher-trainer? Are you still motivated by your role and the mission? \\ \bottomrule
\end{tabular}
\end{table}

\begin{landscape}
\begin{table}[h]
\centering
\caption{Teacher surveys distributed at the end of each PD session. Please note that the 7-Point Likert items (denoted 7Pt Likert) range from -3 (completely disagree) to +3 (completely agree). Supporting references are provided per dimension with Cronbach's alpha for scales with 7 Point Likert items. }
\label{tab:teacher_surveys}
\footnotesize
\begin{tabular}{lllp{9cm}lllll}
\toprule
\multirow{2}{*}{Dimension} & \multirow{2}{*}{Topic} & \multirow{2}{*}{Format} & \multirow{2}{*}{Question} & \multicolumn{4}{c}{PD-session} \\
 &  &  &  & 1 & 2 & 3 & 4 & 5\\ \midrule
 & Grades taught & Checkboxes & Could you please indicate the grade(s) you teach? [1-8, none, other]  & x & x & x & x & x \\
 & School & Checkbox & Which school do you teach in? [list of schools] & x & x & x & x & x \\  \midrule

PD-program evaluation & Interest & 7Pt Likert & The content of the training session was rich and interesting & x & x & x & x & x \\
\ifdefined\Anonymous \citet{anonymous_el-hamamsy_computer_2021}\else \citet{el-hamamsy_computer_2021}\fi & Difficulty & 7Pt Likert & The content of the training session was adapted in terms of difficulty & x & x & x & x & x \\
Cronbach's $\alpha = 0.878$ & Content & 7Pt Likert & I enjoyed the workshops  & x & x & x & x \\
 & Trainers & 7Pt Likert & I appreciated the trainers & x & x & x & x & x \\
 & Exchanges & 7Pt Likert & I enjoyed the exchanges with the other participants & x & x & x & x \\ \midrule

PD workshop evaluation & Interest & 7Pt Likert & I found the workshop interesting & x & x & x & x & x \\
(repeated for each workshop) & Utility for teachers & 7Pt Likert & I found the workshop useful for my teaching & x & x & x & x & x \\
\ifdefined\Anonymous \citet{anonymous_el-hamamsy_tacs_2022} \else \citet{el-hamamsy_tacs_2022}\fi & Utility for students & 7Pt Likert & I found the workshop useful for my students' learning & x & x & x & x \\
Cronbach's $\alpha = 0.906$  & Self-efficacy & 7Pt Likert & I feel confident I can lead this workshop with my students  & x & x & x & x & x\\\midrule

Autonomous motivation &  & & I will implement the Digital education activities seen today with my students because:  &  &  &  &  \\
\citet{angot_dynamique_2013} & Intrinsic motivation 1 &  7Pt Likert & They are interesting to do with my class  & x & x & x & x \\
\citet{gagne_motivation_2010} & Identified regulation 1 &  7Pt Likert & Teaching digital education is important  & x &  &  &  \\
Cronbach's $\alpha = 0.705$ & External regulation 1 &  7Pt Likert & I feel I am supposed to do it  & x &  &  &  \\
 & Intrinsic motivation 2 &   7Pt Likert & It will be interesting to teach & x &  &  &  \\
 & Identified regulation 2 &  7Pt Likert & Digital education is useful for my students  & x & x & x & x \\
 & External regulation 2 &  7Pt Likert & I don't have much choice & x & x & x & x \\
 & Introjected regulation 1 &  7Pt Likert & I want to show others that I can do it & x & x & x & x \\
 & Introjected regulation 2 &  7Pt Likert & I would feel a bit guilty if I did nothing & x &  &  &  \\
 &  &  & At the moment I do not intend to integrate digital education into my teaching & x & x & x & x \\\midrule

Support in the schools &  &  7Pt Likert & I know who to turn to in case of difficulties (pedagogical or technical) related to digital education &  & x & x & x \\ \midrule

Adoption of the & Adoption - binary & Checkboxes & Since the last training session, I taught the following activities [list of activities seen in the previous PD-sessions] & & x & x & x & x \\
pedagogical content & Adoption - duration & Numeric & For each activity I taught, I did {[}X{]} periods with my students  & & x & x & x \\
 & Adoption - support &  & To teach these activities I:  & & x & x & x \\
 & (Cronbach's $\alpha = 0.706$) & Frequency & Sought help from the instructional coaches & & x & x & x \\
 &  & Frequency & Worked with colleagues ahead of time to prepare the activities & & x & x & x \\
 &  & Frequency & Co-taught the activity & & x & x & x \\
 & Adoption - classroom management & Frequency & Taught half of the class at a time & & x & x & x \\
\midrule

 Teacher-trainer model &  &  & Being trained by teacher-trainers  \\
 evaluation & Equivalent & Binary & Is the same as being trained by other adult educators (e.g. the experts from the university of teacher education) & &  &  & x &  \\
 & Current teaching experience & Binary & Is an advantage because teacher-trainers have links to practice &  &  &  & x &  \\
 & From the field & Binary & Is an advantage because teacher-trainers understand my reality / what I experience &  &  &  & x &  \\
 & Lack adult-trainer experience & Binary & Is a disadvantage because teacher-trainers have less experience as adult trainers than other trainers &  &  &  & x &  \\
 & Lack discipline-specific competences & Binary & Is a disadvantage because teacher-trainers have less knowledge of the subject matter than other trainers &  &  &  & x &  \\
 & Lack discipline-specific competences & Binary & Is an advantage or disadvantage for another reason (please specify) &  &  &  & x &  \\
 \bottomrule
\end{tabular}
\end{table}
\end{landscape}

\begin{table*}[!h]
\centering
\caption{Digital Education deployment program number of respondents per PD-session according to their declared grade.
}
\label{tab:deployment_demographics}
\begin{tabular}{l|cccc}
\toprule
\textbf{PD Sessions} & \multicolumn{4}{c}{\textbf{Number of Survey Responses By Grade}} \\
 & \textbf{Grade 1-2} & \textbf{Grade 3-4} & \textbf{Other} & \textbf{Total} \\ \midrule
{Session 1 (March. 2021)} & 255 &  280 &   105 &   640 \\
{Session 2 (September 2021)} & 224 &  266 &   100 &   590 \\
{Session 3 (November 2021)} & 211 &  251 &    88 &   550 \\
{Session 4 (April 2022)} & 218 &  259 &    82 &   559 \\
{Session 5 (September 2022)} & 226 &  239 &    99 &   564 \\
{Session 6 (January 2023)} & - &  - &  - &  - \\ \bottomrule

\end{tabular}%
\end{table*}

\begin{table*}[!h]
\centering
\caption{Digital Education pilot program number of respondents per PD-session according to their declared grade.
}
\label{tab:pilot_demographics}
\begin{tabular}{l|cccc}
\toprule
\textbf{PD Sessions} & \multicolumn{4}{c}{\textbf{Number of Survey Responses By Grade}} \\
 & \textbf{Grade 1-2} & \textbf{Grade 3-4} & \textbf{Other} & \textbf{Total} \\ \midrule
{Session 1 (Oct. 2018)} &  85 &   96 &   109 &   290 \\
{Session 2 (Nov. 2018)} & 130 &  136 &    51 &   317 \\
{Session 3 (Mar. 2019)} &  122 &  134 &    45 &   301 \\
{Session 4 (Apr. 2019)} & 65 &  109 &    35 &   209 \\
{Session 5 (Oct. 2019)} & 142 &  149 &    18 &   309 \\
{Session 6 (Dec. 2019)} & 144 &  154 &    19 &   317 \\
{Session 7 (Mar. 2020)} & 129 &  146 &    19 &   294 \\
{Session 8 (Oct. 2021 - delayed due to COVID)} & 157 &  167 &    24 &   348 \\ \bottomrule
\end{tabular}%
\end{table*}

\subsection{Analysis Methodology}
\label{sec:analysis_methodology}

\subsubsection{Qualitative data analysis from the focus groups and journals}

The reflexive journals and focus groups were analysed inductively to align with the unstructured format employed for the reflexive journals. The objective was to avoid constraining the analyses to a fixed set of hypotheses and ensure that all the themes of relevance to the teacher-trainers were considered. Indeed, the objective was to employ a phenomenological approach to analysing the data by placing the emphasis on the teacher-trainers perception of their lived experience \citep{smith_doing_1999}.
To that effect, the data was transcribed and read by two researchers to familiarise with the corpus \citep{braun_using_2006}. The two researchers then employed a descriptive coding approach to ``summari[se] in a word or short phrase – most often as a noun - the basic topic of a passage of qualitative data'' \citep{saldana_coding_2009} which is considered useful for ``studies with a wide variety of data forms (e.g., interview transcripts, field notes, journals, documents, diaries, correspondence, artefacts, video)'' \citep{saldana_coding_2009}, all the while considering that a passage may be attributed multiple codes. This process was done in several stages. First, the researchers iteratively created the coding manual which included descriptions of the codes and both positive and negative examples. After the researchers verified that the codebook was complete, they proceeded to practice coding a subset of the corpus before coding the full corpus. They then employed a ``reconcile difference via consensus'' approach and achieved 99.7\% agreement on the fully coded corpus \citep{syed_guidelines_2015}, with a kappa statistic of inter-rater reliability \citep{mchugh_interrater_2012} of 0.86 which is considered excellent by \citet{bakeman_observing_1997}.

The codes were then mapped using a focused coding approach \citep{saldana_coding_2009} into the following themes to align with the explicit characteristics of the adapted cascade model:

\begin{enumerate}
    \item Teacher-trainer recruitment criteria: being motivated to be teacher-trainers, being motivated towards Digital Education, having current experience as teachers, being in proximity with participants (teachers)

    \item Teacher-trainers' PD characteristics: being in proximity with the experts, quality and difficulties of the trainer-PD, being equipped as an adult trainer, having (in)sufficient appropriation time, having a view on the whole teacher-PD

    \item Support provided by stakeholders in the curricular reform framework: training in pairs of teacher-trainers, having a relationship with the partner teacher-trainer, exchanging among teacher-trainers, collaborating with instructional coaches and other stakeholders in the schools

    \item Teacher-PD deployment: quality and difficulties of the teacher-PD program, managing time, managing uncertainty, managing heterogeneous groups, managing infrastructure, and managing technical problems
\end{enumerate}

For all these categories we considered the impact on teacher-trainers' self-efficacy (feeling legitimate, not being experts, posture as an adult trainer, feeling of self-efficacy), impact on controllability (adapting the content, having a say in the teachers' PD) and motivation, all known limitations of cascade models.

\subsubsection{Quantitative data from the teacher surveys}

The teacher-survey data was analysed descriptively in a first stage to obtain an overview of teachers' perception of the PD-program, pedagogical content and their integration of the new resources into their practices (i.e. adoption). When comparing dependent variables that only include deployment data, the comparison is done using ANOVA. However, then comparing data from the deployment and the piloting phase, as the most piloting data was acquired on a 4-Point Likert scale, the deployment data is first converted to a 4-Point scale when needed. The two datasets are then compared using Kruskal Wallis' non-parametric one-way ANOVA to account for the fact that data on a 4-Point scale is non-normally distributed. In both cases, the minimum effect size (Cohen's D) that can be detected to obtain a statistical power of 0.8, with a probability of rejecting the null hypothesis $\alpha=0.05$, is taken into account.
Where the adoption data is concerned, the comparison was established using $\chi^2$'s test of independence.
Finally, to account for the use of multiple statistical tests, Benjamini-Hochberg's p-value correction is applied to reduce the false discovery rate.

\section{Results}

\subsection{RQ1: Efficiency of the deployment model from the teacher-trainers' perspectives}

\definecolor{mcolor}{rgb}{0.4, 0.61, 0.70}
\newcommand{\mquoteEF}[3]{
\noindent
\fcolorbox{mcolor}{mcolor}{\parbox{0.97\linewidth}{
``#1'' (#3 - #2)
}
}
\\
}

In this section, we provide insight into the teacher-trainers' perception of the adapted cascade model and in particular in relation to the recruitment criteria (see section \ref{sec:perception_recruitment}), the trainers' PD (see section \ref{sec:perception_trainerPD}), the support provided by key stakeholders in the schools (see section \ref{sec:perception_relationships}) and finally the teacher-PD program (see section \ref{sec:perception_teacherPD}). Please note that we employed pseudonyms to prevent the teacher-trainers' being identified in the quotes.

\subsubsection{Teacher-trainer recruitment criteria}
\label{sec:perception_recruitment}

\paragraph{(A) The teacher-trainers are motivated by Digital Education and training their peers but there are issues of recognition.}
Teacher-trainers report that they ``continue to discover things every day'' (Daniel), something which they enjoy and which
contributes to their motivation to be Digital Education teacher-trainers. Being a teacher-trainer is considered to be a next step in their careers which gives them the possibility to maintain their primary activity as teachers in the field. Several ``are even increasing their working load'' (Charlotte) as teacher-trainers by also participating in the second phase of deployment for grades 1-4 (8/14) which began in parallel with the first.

        \mquoteEF{I've got 30 years of teaching behind me and it's just great to be able to do something other than talking to children and doing something that's motivating. It's really interesting, you talk to your peers, you're equal, it's really great.
        }{Emma}{FG4}

        \mquoteEF{I am still happy to be in the field. I wouldn't like to be just a teacher-trainer.
        }{Charlotte}{FG4}

The adapted cascade model therefore appears to promote teacher-trainers' motivation, but there are issues of recognition. These issues include the fact that (i) they do not have an official title within the department of education, and are therefore just perceived as teachers by school leaders, and (ii) is the same as a ``regular'' teacher's salary and therefore lower than the salary of other trainers (e.g. expert trainers) in the region.

        \mquoteEF{We are taken less seriously by the school leaders than those at the university of teacher education. We have a lot of demands regarding the material for the day, we ask them to put their teachers on leave and replace them, and we don't have a title or recognition like other trainers have. We're only teacher-trainers, we're teachers.
        }{Lucie}{FG4}

        \mquoteEF{There are people who have the same job as us and are paid more than me. Sometimes I think that if its to get paid the same, I'm better off in class. What keeps me here is that I believe in the project, I want to be part of it, I'm attracted by it.
        }{James}{FG4}

\paragraph{(B) The teacher-trainers are teachers and are therefore in close proximity with teachers which contributes to their legitimacy and self-efficacy.}

The fact that the PD-sessions are provided by trainers who are active teachers is considered key to getting teachers' acceptance. When presenting themselves ``as teachers like them, it makes a big difference'' (Michael). The teacher-trainers are ``on equal footing with the teachers'' (Lucas).  By ``doing the same thing every day as teachers, [teacher-trainers] are more legitimate than someone who doesn't set foot in the classroom'' (Lucie). The teachers ``really felt like [the teacher-trainers] were teachers like them and it helped a lot'' (Camilla). The teachers and trainers also did not resort to heavy formalities which ``remove many barriers'' (Sofia) and ``ma[de] it much easier'' (Jennifer).

Several PD-characteristics also appear to contribute to the proximity between the teachers and teacher-trainers. For instance, delivering the PD-sessions in the schools, ``the fact that we go there, we go to them,'' (Jennifer), and that many of the workshops are done in small groups helped teachers and teacher-trainers ``get to know each other better'' (Sofia) and ``make a real connection with everyone'' (Charlotte) which ``is a privilege'' (Unknown).
``The teachers have more confidence in [the teacher-trainers]'' (Sofia), and appreciated that the teacher-trainers were ``benevolent'' (William), and ``there for them which reassured them a lot'' (Sofia). The result is less teacher-reticence towards Digital Education and teachers asking ``but are we going to see you next time?'' (Sarah).

        \mquoteEF{There's a natural closeness in small groups. It's true that even people who are not motivated, who are not interested in technology at the beginning, when we come next to them, sit down and say: now we're going, well, they won't say no!}{Michael}{FG2}

\paragraph{(C) The teacher-trainers are current teachers, and therefore able to test the content in their classrooms and speak from experience which contributes to their legitimacy and self-efficacy compared to expert trainers.} The teacher-trainers consider that it's ``a richness that [they] are teachers'' (Emma), ``that [they] do the same job as [teachers] every day'' (Charlotte), a considerable difference from most expert trainers. Being a teacher means that teacher-trainers ``can more legitimately give a PD than somebody who has not set food in the classroom, and talk about how to implement things in practice'' (Charlotte). This experience contributes to the teacher-PD ``being better received and heard in practice'' (James) by the teachers. Indeed, teachers have an inherent negative bias towards experts who come in and ``have not been in the class for 15 years'' (James). Teacher-trainers are able to ``talk about their experience'' (Emma), ``illustrate with anecdotes, say yesterday afternoon I did this with the grade 2 students, I know what you're talking about'' (James). ``It's not like those from the university of teacher education saying there's this and that and things are fine. Teachers at that point are often doubtful that it's working with their particular students'' (Sarah). When teacher-trainers ``introduce themselves and [teachers] see that [they] are in classes, [...] just that [and] poof, [the teachers] are present, they're with [the teacher-trainers]. They say to themselves, there is something concrete that is going to happen'' (Emma).

        \mquoteEF{I think it's the mindset [teachers] come to the PD with too. It's not a university of teacher education course, it changes their mindset. There is also the fact that it will be more down to earth and not just theory that you don't understand. They come with fewer preconceptions about the course.
        }{Lucie}{FG4}

The teacher-trainers being able to reflect on their experience ``with students, are really concrete elements that help a lot'' (Isabella). It contributes to teacher-trainers ``being credible and teachers feeling they are close to [them] because [they] are in the same terrain'' (Lucas). The teacher-trainers ``know their problems, have the same objectives, have had the same meetings'' (Charlotte). One relevant element is also to ``have already gone through the training, to talk about things [they] have already tested out in the field'' (Charlotte), to be legitimate and credible when delivering the PD. It is important that teacher-trainers ``share [their] failures in the classroom as well'' (Sarah).

        \mquoteEF{I've been part of a pilot school for two years, I'm an instructional coach. It's an advantage, it really gives legitimacy to share my experience and say: in my establishment it's done like that.
        }{Isabella}{FG2}

Indeed, in cases where the PD-session included elements ``that [they] had not experienced in the classroom. [The teacher-trainers] felt that [they] were only bringing theory and that [they were] not very credible'' (Charlotte). This aligns with the reflection made by several teacher-trainers who asked themselves ``how can I train teachers to use certain tools if I have never done them before myself?'' (Camilla) for the pedagogical activities that they themselves had not yet tested out. The teacher-trainers consider that they need to ``come with [their] own experience'' (Camilla) because ``it's easier to explain things when [they] have lived them'' (Jennifer).

\subsubsection{Perception of the teacher-trainers' PD}
\label{sec:perception_trainerPD}

The teacher-trainers consider that they were well trained by the two experts who interacted with them weekly and ``had a training that was quite advanced and help them go in depth on many points'' (Daniel). The experts are globally perceived as ``very at ease and benevolent, which is essential. They both mastered their respective fields very well and formed a very complementary pair'' (William). The experts were also from the field and promoted an open and ``frank'' environment (Sarah). They ``listened to the teacher-trainer's requests'' (William), ``provided good feedback'' (Emma), ``trusted [the teacher-trainers] and were very transparent'' (Sarah), which helped prepare the teacher-trainers for their new role.

The teacher-trainers nonetheless mentioned at multiple occasions that they had a considerable amount of elements that they needed to master. ``There are things at the level of the training that go very quick and that [they] need to appropriate, which takes time'' (Daniel). Although they were ``initially worried about being able to master all the content and be credible [...] in a short lapse of time, and being "experts" compared to the teachers [they] were going to train'' (Sofia), the teacher-trainers felt that they ``had the time to appropriate the PD. It was necessary and reassuring'' (Sofia). They ``felt extremely lucky to have a year to prepare, without which [they] would have felt ill at-ease because [they] would not have mastered the content'' (Lucie).
They generally consider that thanks to the weekly sessions they had over approximately 6 months with the experts before the first teacher-PD session ``[they] were extremely well prepared'' (Lucas) to deliver the PD-sessions. They were therefore ``serene when going into the PD-sessions'' (Michael). They did however worry about certain cases (e.g. robotics) ``where [the trainers] are not much ahead of the teachers'' (Jennifer), and do not know much more than the teachers. In these cases, the teacher-trainers would like to have additional time dedicated to them in their own PD. Nonetheless, they consider that as time progresses, they are more efficient and ``will work more quickly and appropriate [the new content] more quickly'' (Olivia).

As they become more comfortable in their posture as adult trainers, the teacher-trainers are also less in ``recital-mode'' and better able to adapt their PD to the group and their needs. This is facilitated by the fact that throughout their initial PD they acquired a complete view of the PD-program which means that they are better able to judge where and how they can adjust the PD.

        \mquoteEF{What I like a lot is that depending on how you perceive the group, there are things that come up to say to this group. There's this thing that they need you to tell them, that you put forward. Depending on the group, there are slides that move, that get thrown out. }{Jocelyn}{FG1}

        \mquoteEF{Even if we were beginners and it was the first time in training, we still had a global vision of what we had to do. This allowed us to regulate, add and remove. We know that if we remove something, it's not serious because we know the globality of the training, of what we have to bring teachers. I really have the impression of mastering the tool, of coming there with credibility in relation to the people by saying: I know where I'm going, I know why I'm here and I know that I have to take you there in two and a half years, and I know that I can. }{Camilla}{FG2}

Nonetheless, the experts are still perceived as being more efficient in delivering the PD-program than the teacher-trainers, and more knowledgeable about Digital Education, notably because they conceived and already piloted the content and PD-program.

        \mquoteEF{[The experts] have a better overview as they have designed the training so they know from A to Z how far it goes. They are probably better at creating links. What takes us a long time to do, should be easier and more efficient for them. In addition, they have been giving the training for a long time.
        }{Camilla}{FG4}

        \mquoteEF{That's when you see that the people who designed the PD-program are hyper-experts, who speak quickly, who know exactly what they want to say.  Whereas we have to take the time to get to grips with it all. We also adapt our flow to the understanding of the people in front of us. }{Olivia}{FG4}

The teacher-trainers also consider that they would have liked ``more PD on how to be an adult trainer [...] because [they] are in front of children all day, not adults, and don't speak the same way to both. [They] think that [they] are still not trained enough and would have liked to be trained more in this area'' (Charlotte).

        \mquoteEF{I didn't feel legitimate in relation to adult training. And I found, that the adult-training we had was short [...] for people who had never trained before. We learned a lot on the job. But well, it worked. The important thing is that [teachers] can ask questions and I can answer them.
        }{Ava}{FG4}

\subsubsection{Relationships and support within the curricular reform framework}
\label{sec:perception_relationships}

\paragraph{(A) The teacher-trainers support one another throughout the process, whether in pairs or as a whole group}

Being placed in pairs to train-teachers is viewed positively by the teacher-trainers. Although it was not the case in the pilot program, the pairs were fixed for the deployment phase, following feedback obtained through the reflexive journals during the first weeks of their trainer-PD. This gave the teacher-trainers the ``time to get to know each other'' (Sofia), to know ``how to work and build something together in the long term'' (Camilla), and become an invaluable support throughout the process. Being in pairs that know each other well not only helps ``with all the things you need to manage at the same time, but also to relay one another when needed'' (William). They ``know the strengths of [their] partner, the weaknesses, and how to bounce back'' (Emma).  Over time they ``are better able to coordinate'' (Sofia). The teacher-trainers therefore consider that ``there is a great deal of added value in having the same pairings. If you were to rotate, you would lose a lot of training quality'' (James) and it would ``require doing the work over again multiple times'' (Emma).

        \mquoteEF{It's very important [to train in pairs]. If I was on my own to manage the day, it's a lot, there are so many things to do.
        }{Lucie}{FG4}

        \mquoteEF{When you're training in pairs, it also allows you to have someone who listens and who can bounce back. It's more than just having someone who listens a bit. We are constantly ping ponging.
        }{Emma}{FG4}

The teacher-trainers also support each other as a whole team. For instance, while on the days where they are delivering the PD-sessions, they ask each other questions in the instant messaging group ``and have an answer in under ten minutes that [they] can show to teachers'' (Sarah),  they are ``super reactive [... and] know that there is always somebody who will be able to give an answer. It's very precious'' (Camilla).

        \mquoteEF{This year [of training] has been really beneficial for the team because we have got to know each other. We've been through things together, we've had our difficulties together and I think we have a strong bond. I think we have a great team.
        }{Olivia}{FG4}

\paragraph{(B) The teacher-trainers create important links with teachers and instructional coaches in schools}

In addition to the links among each other, it is also important to have a positive relationship with teachers and to maintain these overtime. These relationships with teachers are beneficial and quite ``innovative'' (Charlotte) in this PD-program as the teacher-trainers are practising teachers in the region (see section \ref{sec:perception_recruitment}), although it may be delicate when it comes to being a teacher-trainer in your own school because it can ``be a bit more personal [... so] you also have to protect yourself a little bit, to take a step back from the affect'' (Olivia).

        \mquoteEF{Because we have created these links with teachers we know them. They know us, we eat lunch with them. It's beneficial to go to the same places for the teachers for whom it goes well, but if the teachers don't feel it, you have to change pairs.}{Lucas}{FG4}

The other key stakeholders in this ecosystem are the instructional coaches who are ``like janitors in colleges or cooks in camps. If you don't have them in your pocket, you can't do anything you want'' (Emma). ``Creating links with the instructional coaches is great because when [the teacher-trainers] go back, [they] already have something that is established with the people in the school'' (Sarah). The teacher-trainers ``rely on the instructional coaches [...] it's very important that they are in schools. It's essential to have them, that they are identified in the schools. They carry the [Digital Education] project on their shoulders when we leave'' (Camilla). Having these instructional coaches ``is a chance [...], without them what [the teacher-trainers] are doing would be useless because the teachers would not really adopt'' (Daniel). Indeed, the instructional coaches play a critical role in the way the teacher-PD content is applied between sessions.

        \mquoteEF{The instructional coach doesn't make it work for each individual (that depends more on teachers' personal motivation) but more on the level of the whole school so that everyone is involved and tries to do something. The coach makes sure that all the students will have done some Digital Education at some point, but you know that there will be students who will have done more than others. }{Charlotte}{FG4}

However, these instructional coaches also need to be prepared and trained, to understand their role, and contribute to the specifics of the organisation of the session in the schools (e.g. rooms, access to the wifi, material, infrastructure), in addition to supporting teachers in their implementation of the discipline. Although it is generally the case in the project as all the instructional coaches are trained by the experts prior or in parallel to the deployment, there are cases where the ``instructional coach is new and doesn't know anything'' (Lucas),  which complicates things. Ideally there would even be protocols in place in all schools to facilitate the ``organisation of the day to book the rooms, get the tables, material, beamer...'' (Lucie) and avoid the disparities observed between schools. Nonetheless, ``it does not change much, although it more pleasant to go where the instructional coach is nice, organises things a bit ahead of time. But it doesn't change much for the course we give'' (Lucas).

\subsubsection{Perception of the teachers' PD}
\label{sec:perception_teacherPD}

\paragraph{(A) The teacher-PD was already conceived prior to recruiting teacher-trainers, which contributes to the teacher-trainers feeling they lack ownership.}

The teacher-trainers generally perceive the PD-program as being ``of high quality'' (Michael), ``logical and well done [...] which is reassuring'' (Emma). The content is ``down to earth and not just theory that you don't understand. Teachers come to the training with less preconceived notions about the PD-program'' (Lucie). However, the teacher-trainers ``provide PD-sessions that were designed by other trainers'' (Daniel), they are told what to do, and feel like ``spokespersons'' (Sofia). They appreciate ``the space given to speak during the training, and being able to review the teacher-trainer PD'' (Isabella) and feel that experts ``adapt the form but not the content of the PD-sessions'' (Daniel), e.g. the global outline of the PD-program and the selection of PD-activities that were piloted prior to their recruitment. The result is that they feel ``they can't [always] afford to eliminate [a given] part'' (Daniel), particularly since they ``have to provide a common culture in the region'' (Charlotte), which although positive to avoid content dilution, shows that teacher-trainers feel they lack ownership and control over the global decisions related to the teacher-PD program's content.

        \mquoteEF{I don't allow myself to eliminate completely certain parts of the training because if they're there, it's because they were designed to be there, but I do add a little bit of my own personal touch}{Daniel}{FG4}

The teacher-trainers nonetheless recognise that the specifics of a given PD-session are adapted to their feedback as they deploy program, and that they are able to make adaptations to the PD-sessions (see paragraph C below).

        \mquoteEF{Every time we came with remarks, they were taken into account. [The experts] put in a lot of effort. So we are heard and for a lot of other things too, we were able to redo things, we were able to adjust them. There is a real collaboration and it is precious.}{Camilla}{FG4}

\paragraph{(B) The teacher-trainers consider that the teacher PD-sessions generally go well, with the exception of some technical difficulties.} Indeed, there were not many difficulties due to heterogeneous groups. Even ``teachers who knew a lot did not show it off'' (James). Sometimes, teachers who were already familiar with the tools and competencies even ``helped out their colleagues, explained things'' (Camilla). ``Apart from technical elements, there's not much else that went wrong'' (Lucie). Indeed, the main issues they encountered were related to elements of the school infrastructure (e.g. connecting with the beamer, wifi, ...), with conditions which varied greatly across schools.

\paragraph{(C) The teacher-trainers make adaptations to improve a given PD session, but run the risk in certain cases of diluting the content.}\hspace{1pt}

The first type of adaptations are for a specific PD-session. As teacher-trainers repeat the PD-session with different groups of teachers, they become more at ease with the content, adapt ``the keynotes'', ``speech and examples to the group'' (Sarah), and ``the timing'' (Lucie) for instance when ``it's too long and see that the teachers are bored'' (William). The teacher-trainers consider that they ``are clearly improving'' (Camilla), notably when they adapt based on ``the teachers' feedback at the end of a day'' (Charlotte). The result is that the last cohort of teachers experiencing a given PD-session receive better quality training. Nonetheless, it is important to be aware that too many consecutive repetitions of same PD-session can be taxing for the teacher-trainers.

        \mquoteEF{As the days progress, I add information, I create links. In fact, the better I master the material, the better I can create links with things that have already been seen, with things that will come.
        }{Camilla}{FG4}

        \mquoteEF{We also know which questions systematically come back. And then we can also say: we know that this is also something that could be difficult and answer the question before it pops up.
        }{Sofia}{FG4}

        \mquoteEF{I think that indeed, at least for our pair, we evolved between the first and the last session. As we progressed, we told teachers the surveys were important and read the comments of the surveys to adjust each time a little bit more.
        }{Olivia}{FG4}

        \mquoteEF{Ava - Giving the same session 5 times is not bad because we can improve quite a bit and then move on. But giving 14 times a half-day is too long!

        Olivia - What saves us also is that we don't have the same public each time in front of us [...] but again, it's also part of our career, to give the same program 25 years in a row to 25 different groups of students so...
        }{Multiple}{FG4}

There is however a more considerable risk of content dilution when the trainers make adaptations to counterbalance the fact that the PD is on ``a timing that is quite strict'' (Isabella). Indeed, the trainers mention ``having considerable delays which is difficult and means that [they] had to cut things'' (Isabella) in order to adapt. The bigger issue however is when teacher-trainers forgo a (portion of a) workshop, such as ``reducing the theory to spend more time on the practice'' (Michael), or not using one of the tools the teachers were supposed to be introduced to ``to put the accent on [another workshop] so that they practice it rather than just listen passively'' (Sofia). \\

The second type of adaptations are for a group of teachers being followed up by the same pair of teacher-trainers, because they ``start to know them and are able to insist where [teachers] need to so that [teachers] adhere to the program'' (Emma). ``The teachers also progress as the days go by, there is an increase in power'' (Lucie).

        \mquoteEF{I would even say that for a same group we evolve from one time to another because we know them. And then, we know what will go well and what will not. And then, depending on the instructional coach who is present, we know we have to pay more attention to this thing or that thing depending on the context. Depending on these factors we don't approach the day in the same way.
        }{Sofia}{FG4}

Generally the results indicate that the teacher-trainers are attuned to the teachers' needs and adapt what they teach accordingly. This is possible because teacher-trainers ``have some flexibility [in the way they organise the sessions], which is good'' (Sofia) which helps them ``adapt their discourse to the group, give different examples'' (Sarah). Teacher-trainers are able to ``bring [their] personal touch for each group according to their needs and reactions
'' (Daniel).

        \mquoteEF{We also had people who came in complaining and saying: I don't know anything anyway. So we took a lot of time with them. Step by step. And then it went well.
        }{Charlotte}{FG2}

They do however consider that the PD-program can be intense, and that it's better if the PD is spread out over time to facilitate teachers' appropriation of the content.

\paragraph{(D) The teacher's PD is generally well received by teachers. }
Despite the challenges, the teacher-trainers report that the teachers are generally satisfied with the PD-program and complain little(coherently with the results of the teacher surveys in section
\ref{sec:deployment_teacher_PD}). Even ``those who
complain are still motivated'' (Camilla). The teacher-trainers ``were pleasantly surprised because they were prepared for 5\% of grumblers [...] and in the end, [they] hardly saw them'' (Michael).
Typical reluctances however remain towards introducing activities that employ screens with young students, particularly after COVID where ``students already used screens for too long'' (Camilla). Nonetheless it appears that participating in the workshop contributes to attenuating these reactions over time as ``it didn't really come out afterwards'' (Camilla).
Additionally, two extremes of teachers appear to pose a considerable challenge:

\begin{itemize}
    \item Young teachers who ``feel like they know everything'' (Michael) and are ``sure of knowing everything'' (Sofia), partly because ``at the university of teacher education they already have Digital Education modules and have undergone part of the training already'' (Lucie).

    \item ``Teachers who are a year away from retirement, [who] didn't want to get involved, that it never interested them and that they were never going to do it anyway'' (Michael).
\end{itemize}

Finally, one more general issue that arises is the adaptability of the PD-program to specialised teachers (sports, arts and crafts etc...) who are unclear on what the utility of the PD program is for them, despite the fact that ``in craft, sewing and sports classes, you can really find use cases'' (Jennifer). Indeed, one marker of the project is that it is ``important to provide a common culture'' (Daniel) to all teachers, but this requires ``giving [these specialised teachers] more activities that are ready to use in their contexts'' (Charlotte).

\subsection{RQ2: Efficiency of the deployment model from the teachers' perspectives}
\label{sec:deployment_teacher_PD}

\subsubsection{Teachers' perception}

\begin{figure}[h]
    \centering
    \includegraphics[width=\textwidth]{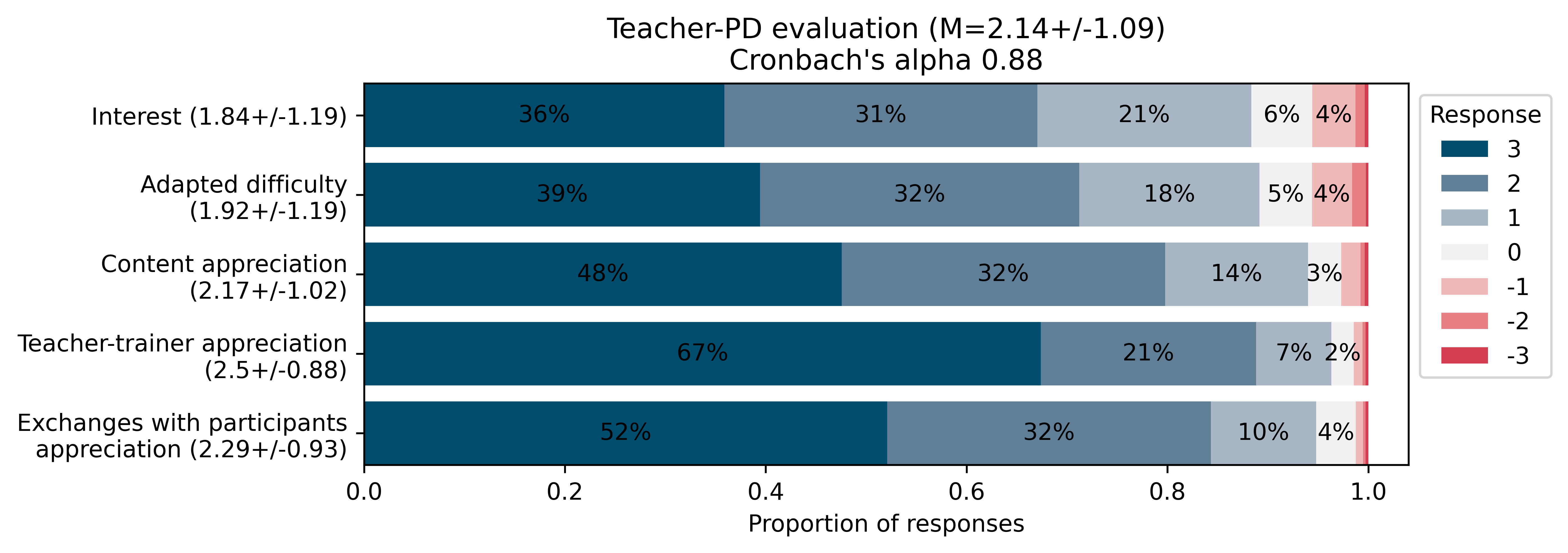}
    \caption{Deployment teacher-PD program evaluation with responses aggregated over the 5 sessions. Cronbach's $\alpha$ is provided as an indicator of the reliability of the scale.}
    \label{fig:TPD_eval}
\end{figure}

The teachers in the deployment phase positively perceive the teacher-PD program according to the global evaluation metrics (see Fig. \ref{fig:TPD_eval}), with no significant differences according to grade ($F(1)<=1.671$, $p>0.19$). Indeed, over all sessions, the PD-program evaluation criteria garnered an average between totally agree and agree on the 7-Point Likert scale. Furthermore, for all evaluation criteria, over 79\% responded positively with at most 5\% of negative responses. These results are significantly better with at least a medium effect size for all 5 categories in the deployment phase with the adapted cascade model compared to the pilot program where the teachers were trained by experts (see Table \ref{tab:TPD_eval_comp}).

\begin{table}[h]
    \caption{Comparison of the evaluation of the teacher-PD program between the piloting phase and the deployment phase using Kruskal Wallis' non parametric one way ANOVA on the data that was converted to a 4 Point Likert scale (between 1 - completely disagree, and 4 - completely agree)}
    \label{tab:TPD_eval_comp}
    \centering
    \begin{tabular}{p{2cm}cccccc}
    \toprule
    {} & Pilot M+/-SD  & Deployment M+/-SD &   Statistic &    p &  Cohen's D &          Comparison \\
    {} & (4-Point Likert) & (4-Point Likert) \\
    \midrule
    Interest      &    3.16+/-1.01 &         3.42+/-0.59 &         38.8 &  0.0 &      -0.31 &  Deployment $>$ Pilot \\
    Adapted difficulty             &    3.12+/-1.00 &         3.46+/-0.60 &        130.4 &  0.0 &      -0.42 &  Deployment $>$ Pilot \\
    Content appreciation   &    3.14+/-1.01 &         3.59+/-0.51 &        234.0 &  0.0 &      -0.56 &  Deployment $>$ Pilot \\
    Teacher-trainer appreciation &    3.18+/-0.82 &         3.75+/-0.44 &        611.5 &  0.0 &      -0.86 &  Deployment $>$ Pilot \\
    Exchanges with participants appreciation  &    1.54+/-1.22 &         3.65+/-0.47 &        901.2 &  0.0 &      -2.29 &  Deployment $>$ Pilot \\
    \bottomrule
    \end{tabular}

\end{table}

\begin{figure}[h]
    \centering
    \includegraphics[width=\textwidth]{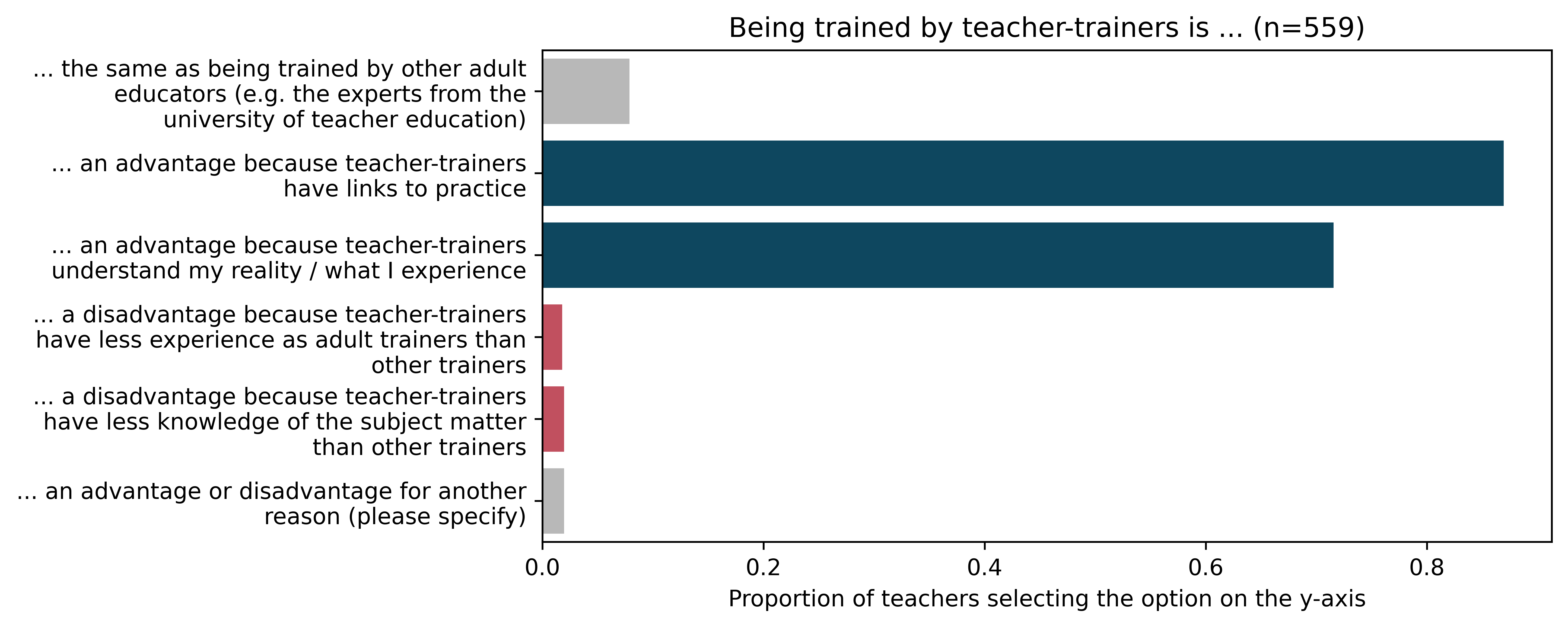}
    \caption{Deployment teachers' perception of the teacher-trainer deployment modality}
    \label{fig:teacher_trainer_eval}
\end{figure}

The teacher-trainers are particularly highly rated with less than 2\% of negative responses over all sessions. Coherently with the teacher-trainers' opinions expressed during the focus groups, the teachers prefer having teacher-trainers rather than experts to a large extent (see Fig. \ref{fig:teacher_trainer_eval}). Indeed, 87\% viewed the teacher-trainer model as an advantage because the teacher-trainers have links to practice, and 72\% because they understand their reality. Just 2\% report that it is an inconvenience that the teacher-trainers have less adult-training experience, and 2\% that they have less Digital education competences that expert trainers.

\subsubsection{Teachers' adoption of pedagogical content into their practice}

\begin{figure}[h]
    \centering
    \includegraphics[width=\textwidth]{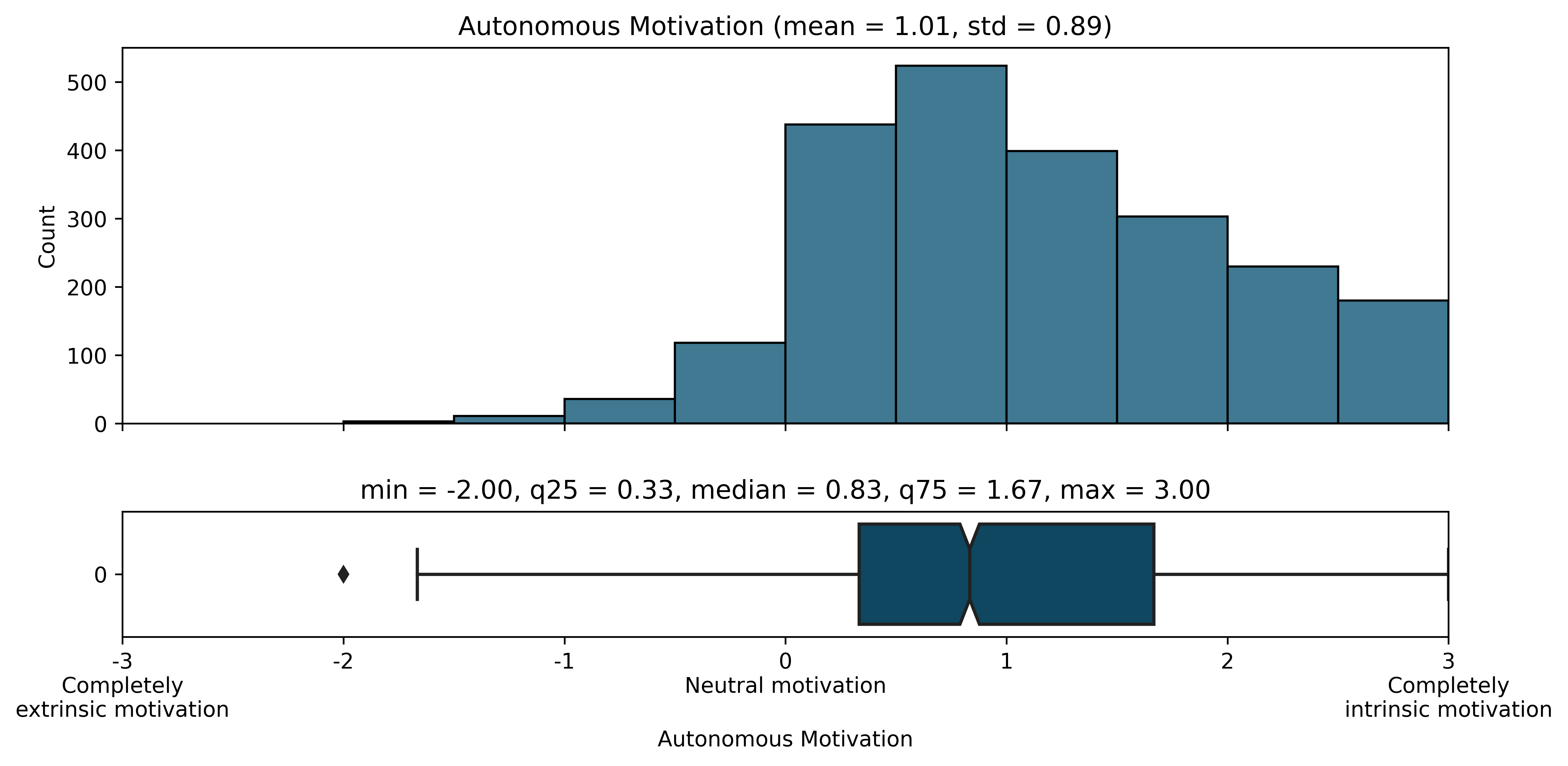}
    \caption{Deployment teachers' autonomous motivation to teach the Digital Education pedagogical content computed using the Relative Autonomy Index \citep{grolnick1989parent} which combined the Autonomous motivation sub-scales into one dimension according to $AM=(2\times IM +1\times IdR-1\times InR-2\times ER)/6$ where AM is the overall autonomous motivation, IM the intrinsic motivation items, IdR the identified regulation items, InR the introjected regulation items and ER the external regulation items.  An autonomous motivation of -3 is completely externally regulated, an autonomous motivation of 0 is neutral, and an autonomous motivation of +3 is completely intrinsic.}
    \label{fig:AM}
\end{figure}

\begin{figure}[h]
    \centering
    \includegraphics[width=\textwidth]{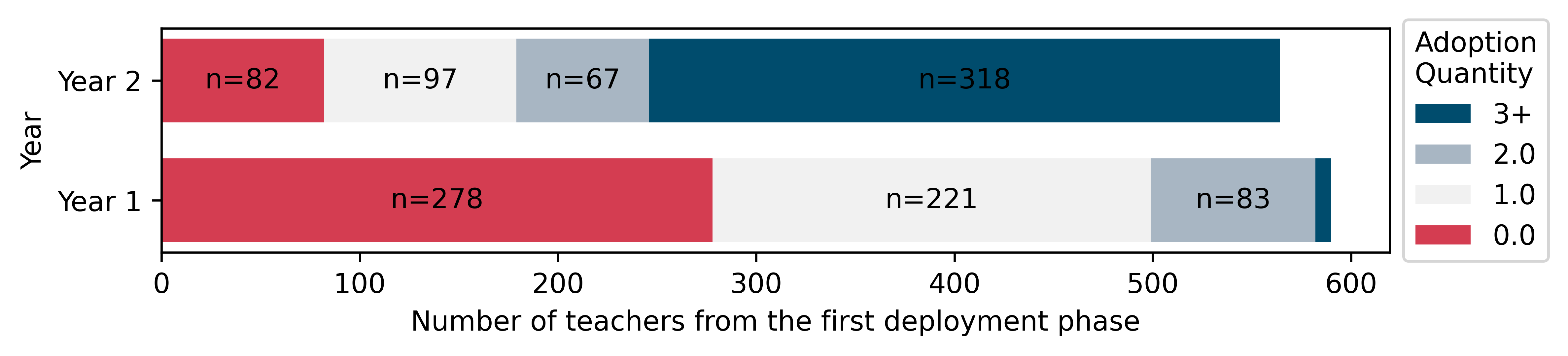}
    \caption{Number of Digital Education pedagogical activities adopted by the teachers in the deployment phase in the first and second year (i.e. adoption quantity, $\mu_{year1}=2.57\pm1.8$, $\mu_{year2}=3.0\pm2.5$). Please note that the year 1 adoption rates correspond to what the teachers taught between the first and second PD session, while the year 2 adoption rates correspond to what the teachers taught between the second and fourth teacher-PD sessions.}
    \label{fig:Adoption}
\end{figure}

The teachers are globally autonomously motivated to teach the pedagogical content seen in the PD-sessions, with over 85\% being on the positive end of the spectrum (see Fig. \ref{fig:AM}). The result is that 75\% of teachers adopt at least one Digital Education pedagogical activity in the second year of the program (with an average $\mu_{year2}=3.0\pm2.5$ of number of activities adopted, see Fig. \ref{fig:Adoption}). $\chi^2$'s test of independence further indicates that there are no significant differences between adoption patterns between the pilot and deployment phases ($\chi^2(16)=17.5$, $p=0.3566$). Teachers in the deployment phase therefore adopt to similar extents as those in the pilot program, further demonstrating the efficacy of the deployment model.

\begin{figure}[h]
    \centering
    \includegraphics[width=\textwidth]{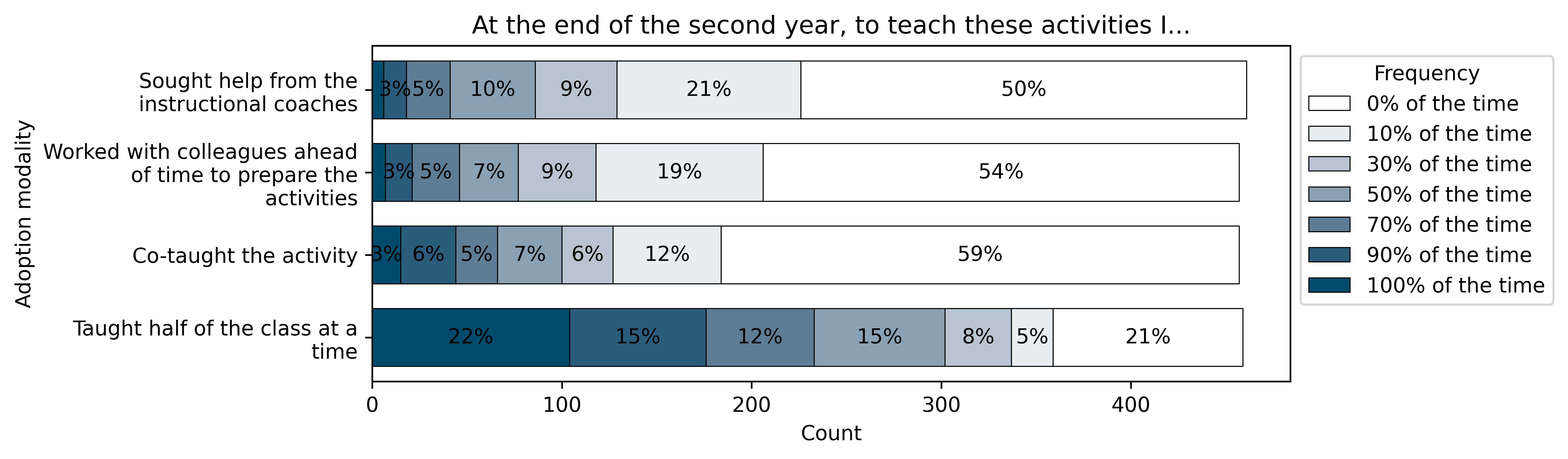}
    \caption{Adoption modality employed by the teachers to teach the Digital Education pedagogical content at the end of the second year of deployment.}
    \label{fig:Adoption_modality}
\end{figure}

Finally, although we did not inquire into how the instructional coaches and the community of practice within each school contributed to engaging teachers to teach the discipline, we note that at the end of the second year, 50\% solicited help from the instructional coach, 46\% worked with colleagues ahead of time, and 41\% co-taught at least one activity (see Fig. \ref{fig:Adoption_modality}). These findings therefore confirm the importance of having the support of instructional coaches, as well as a community of practice in the schools.

\subsection{Methodological Limitations}

With respect to the teacher-trainer data, we only considered their perception. Although for most metrics (e.g. sufficient support, planning, motivation, self-efficacy) this is not an issue, we did not have access to objective measures when it comes to content dilution. It would indeed be important to gain insight into how teacher-trainers deliver the PD-program and to what extent there is implementation fidelity, i.e. ``the degree to which an intervention or programme is delivered as intended'' \citep{carroll_conceptual_2007}. Only then will we be able to better align what is intended by the curriculum with what the students learn in classrooms \citep{van_den_akker_curriculum_2003}. To that effect, a study is presently under way to establish how the teacher-trainers are delivering the PD-program to understand not only whether the PD-program sessions are delivered as intended, but also why they are making adaptations \citep{century_implementation_2016} and their consequence in terms of PD-quality and content dilution in the cascade.  \\

With respect to the teacher-survey data, there are two main limitations. The first is the fact that the evaluation framework evolved between the pilot program and the deployment, and therefore they did not always include the same metrics in order to do a complete comparison, or employ the same scale (i.e. shifting from a 4 point to a 7 point scale). The second was the challenge of reliably tracking teachers over the course of their PD-sessions to link their perception with their adoption in subsequent sessions and in the long term. Indeed, due to inconsistent responses in the pseudonym construction questions, every additional survey contributed to a loss of 30\% of the consistently tracked responses.  \\

With respect to the comparison between the deployment and pilot outcomes, there are factors that may contribute to changing teachers perspective on Digital Education and the PD-program which are independent of the deployment's modalities. For instance, the pilot program took place pre-COVID while the deployment is post-COVID. Such an element may have influenced teachers' readiness to adopt Digital Education. From this perspective, it would therefore be interesting to conduct a similar analysis with other levels of schooling where the piloting phase began after the pandemic. Another factor which may play a considerable role in the teachers' improved perception of the PD-program is the iterative refinement that is conducted by the experts both during the piloting and the subsequent deployment phases which contributes to improving the quality of the PD-sessions and associated resources for following cohorts of teachers. Nonetheless, such an element represents a strength of the overall curricular reform framework. \\

Finally, with respect to the overall evaluation framework, we mostly focused on spread and establishing the efficacy in terms of the limitations of cascade models and teachers' perception of the PD-program and adoption. However, we did not consider other elements pertaining to the spread of ``reform norms, beliefs, and principles within schools and districts'' \citep{coburn_rethinking_2003}. In particular it would be important to gain insight into the long term shift in reform ownership and depth of changes in teachers' practices.  \\

\section{Conclusion}

The debate regarding the introduction of Digital Education into formal curricula has shifted from whether to how this introduction should be done. This question is particularly pressing considering the challenges involved with affecting and sustaining changes in teachers' practices, and doing so at scale, which are only exacerbated in the context of Digital Education and particularly at the primary school level. Unfortunately, the literature indicates that this challenge has not yet been effectively addressed, with many countries still lacking a sufficient number of adequately trained teachers to introduce the discipline at all levels of formal education. To that effect, it appears essential to propose a centralised professional development (PD) for all teachers, whether in-service or pre-service. While pre-service teacher education can be handled by universities of teacher education, the more considerable challenge is providing adequate PD to all in-service teachers, particularly when there are a reduced number of experts and a limited time frame to deploy. Cascade models of in-service teacher training therefore appear to be a promising means of addressing these challenges. They rely on experts training trainers who themselves train other trainers until reaching the end of the cascade and the teachers. Unfortunately, cascade models suffer from considerable limitations which have even led to several initiatives reporting failures. These challenges include the fact that they are one-way transmissive of information, do not always sufficiently support or train the trainers to take on their new role and effectively deploy the PD-program, suffer from content dilution, and a misalignment of the PD with teachers' context and needs. \\

In this article we thus consider the case of a Digital Education curricular reform project which considered sustainability and scalability as key outcomes from the start in order to effectively introduce Digital Education for all K-12 students in the region. To that effect, the project is embedded within a Research Practice Partnership framework that sought to pilot the reform and associated PD-program for all levels of schooling prior to large scale deployment. Considering the primary school reform for grades 1-4 which marked the starting point for this project we first validated the curricular reform model and the associated PD-program with 350 grade 1-4 teachers in terms of perception, adoption, \ifdefined\Anonymous \else \citep{el-hamamsy_computer_2021} \fi, sustained adoption \ifdefined\Anonymous \else \citep{el-hamamsy_sustainability_2022} \fi and student-level outcomes \ifdefined\Anonymous \else \citep{el-hamamsy_primary_2023} \fi. The next objective was thus to spread the reform to the entire region, i.e. all 9'000 teachers and 130'000 students. It was therefore essential to propose an adapted cascade model to effectively deploy the reform to the region all the while seeking to address the limitations of cascade models. \\

The contribution of the present study therefore lies in the proposed adapted cascade model. Used to deploy the Digital Education curricular reform and PD-program to an entire administrative region, the adapted cascade model is anchored in recommendations from the literature and has the following characteristics which differ from other cascade models :
\begin{itemize}
    \item having expert trainers who are ex-teachers from the field, who conceived and piloted the PD-program (and are therefore the most credible), train the teacher-trainers and handle the logistics of the deployment
    \item limiting the depth of the cascade model by having these expert trainers be in direct contact with all teacher-trainers who will deploy the PD-program in the region
    \item recruiting teacher-trainers from the region, and ideally the level of schooling targeted, based on their motivation towards Digital Education and adult training, and ensuring that they maintain a teaching position in order to remain linked with the field and be able to test the content that they will disseminate in the PD-program in their own classrooms
    \item having the experts provide a prolonged initial PD to the teacher-trainers to help them acquire the required competences in terms of Digital Education and adult training, all the while remaining aware of the additional costs and time required to do so
    \item having the experts provide a continuing PD to the teacher-trainers throughout the deployment to help them prepare for each PD-session, support them while they are delivering the PD-session, debrief on their experience and adjust the PD-program accordingly
    \item deploying the PD-program to teachers with pairs of teacher-trainers who go to the same schools in order to establish relationships with teachers in the field
    \item employing instructional coaches in the schools where the PD-program will be delivered to collaborate with and support teachers in the implementation of the discipline and help teacher-trainers organise and adapt the school-level teacher-PD to the teachers' needs (and more generally having links between all project partners, i.e. school leaders, instructional coaches, teacher-trainers)
    \item having action research as a core component to monitor the outcomes of the deployment, and provide feedback to teacher-trainers and experts to adapt the teacher-trainer PD and the teacher-PD program
    \item ensuring that teacher-trainers interact with researchers so that they understand the objectives, methodologies and engage in the process
\end{itemize}

To validate the adapted cascade model, we evaluated the outcomes using (i) qualitative teacher-trainer data from the 14 teacher-trainers employed to deploy the PD-program and (ii) quantitative data from 700 teachers who participate in the first deployment phase, which we compared with data from 350 teachers who participated in the pilot phase. \\

The results demonstrate the effectiveness of the adapted cascade model, by having teacher-trainers who are motivated, consider themselves to be legitimate as they tested the content themselves, and are in proximity with the field. Their prolonged trainer-PD and the support of the experts throughout the deployment process are considered key to achieving this objective. The teacher-trainers establish meaningful relationships with teachers and instructional coaches, and adapt their PD-sessions to the teachers needs. The result is highly positive teacher outcomes as teachers rate the PD-program more positively than in the pilot and prefer having teacher-trainers than expert trainers. The teachers are autonomously motivated to teach the discipline, and adopt to similar extents as in the pilot.
These findings combined validate the deployment model and contribute to providing additional recommendations which may further improve the outcomes such as: ensuring that the trainers have more adult-trainer PD, official recognition in their new roles (e.g. salary augmentations, reduced working times, official recognition with a title...), and that they have ownership of the reform and there is sufficient implementation fidelity to further assess the quality of the deployed PD-program.
Indeed, one limit of the present curricular reform framework is that the teacher-PD resources were already piloted and validated with experts which guarantees the coherence with the full reform. However, this process did not include the teacher-trainers who ultimately deploy the teacher-PD at a large-scale, as they were not yet recruited at the time. Other deployment models may therefore consider recruiting teacher-trainers in earlier phases of the reform and including them as co-conceptors of the program so they may acquire a sense of ownership, in addition to improving the co-constructive process by including active teachers in the conception team \ifdefined\Anonymous\citep{anonymous_el-hamamsy_coconstructing_2022}\else\citep{el-hamamsy_coconstructing_2022}\fi . \\

To conclude, in this article we demonstrated that the deployment framework is effective to introduce Digital Education in the region and addresses many known limitations of cascade models, all the while achieving results that are similar, if not better, than in the pilot program in certain regards. Nevertheless, other factors are decisive for the success of such endeavours such as the curricular reform framework which helped set the stage \ifdefined\Anonymous\citep{anonymous_el-hamamsy_computer_2021}\else\citep{el-hamamsy_computer_2021}\fi.
Other endeavours seeking to employ the deployment framework should therefore consider this conjointly with the lessons learnt from the global curricular reform framework \ifdefined\Anonymous \citep{anonymous_el-hamamsy_computer_2021} \else \citep{el-hamamsy_computer_2021} \fi  in order to increase the likelihood of their endeavour succeeding.

\ifdefined\Anonymous

\else

\section*{Acknowledgements}

We would like to thank the participants and members of the different institutions (Department of Education - DEF, the University of Teacher Education – HEP Vaud, the teams of the two universities - EPFL and Unil) for supporting the EduNum project led by the minister of education of the Canton Vaud.
This work was funded by the {the NCCR Robotics, a National Centre of Competence in Research, funded by the Swiss National Science Foundation (grant number 51NF40\_185543)}.
\fi

\section*{Data availability}

The teacher survey data from the deployment and pilot program will be publicly available on Zenodo upon publication  (\citealp{el-hamamsy_scalability_dataset}, doi:10.5281/zenodo.7912941).

\bibliographystyle{apa_}
\bibliography{0-bib}